\documentclass[
 reprint,
superscriptaddress,
 twocolumn,
 amsmath,amssymb,
 aps,
 prb,
]{revtex4-2}

\usepackage[pdftex]{graphicx}
\usepackage{graphicx}
\usepackage{dcolumn}
\usepackage{bm}
\usepackage{physics}
\usepackage{mathrsfs}

\usepackage{color}

\begin{document}

\preprint{APS/123-QED}

\title{Systematic investigation of dynamic nuclear polarization with boron vacancy in hexagonal boron nitride}

\author{Yuki Nakamura}
\author{Shunsuke Nishimura}
\affiliation{ 
Department of Physics, The University of Tokyo, 7-3-1 Hongo, Bunkyo, Tokyo 113-0033, Japan
}%

\author{Takuya Iwasaki}
\affiliation{%
Research Center for Materials Nanoarchitectonics, National Institute for Materials Science, 1-1 Namiki, Tsukuba, Ibaraki 305-0044, Japan
}%

\author{Shu Nakaharai}
\affiliation{%
Department of Electric and Electronic Engineering, Tokyo University of Technology, 1404-1 Katakuramachi, Hachiohji, Tokyo 192-0982, Japan
}%

\author{Shinichi Ogawa}
\author{Yukinori Morita}
\affiliation{%
 National Institute of Advanced Industrial Science and Technology, 1-1-1 Umezono, Tsukuba, Ibaraki 305-8568, Japan
}%

\author{Kenji Watanabe}
\affiliation{%
Research Center for Electronic and Optical Materials, National Institute for Materials Science, 1-1 Namiki, Tsukuba, Ibaraki 305-0044, Japan
}%

\author{Takashi Taniguchi}
\affiliation{%
Research Center for Materials Nanoarchitectonics, National Institute for Materials Science, 1-1 Namiki, Tsukuba, Ibaraki 305-0044, Japan
}%

\author{Kento Sasaki}
\affiliation{ 
Department of Physics, The University of Tokyo, 7-3-1 Hongo, Bunkyo, Tokyo 113-0033, Japan
}%

\author{Kensuke Kobayashi}
\affiliation{ 
Department of Physics, The University of Tokyo, 7-3-1 Hongo, Bunkyo, Tokyo 113-0033, Japan
}%
\affiliation{Institute for Physics of Intelligence, The University of Tokyo, 7-3-1 Hongo, Bunkyo, Tokyo 113-0033, Japan}
\affiliation{%
Trans-scale Quantum Science Institute, The University of Tokyo, 7-3-1 Hongo, Bunkyo, Tokyo 113-0033, Japan}%

\date{\today}

\begin{abstract}
Dynamic nuclear polarization (DNP) using the boron vacancy ($\mathrm{V_B^-}$) in hexagonal boron nitride (hBN) has gained increasing attention. 
Understanding this DNP requires systematically investigating the optically detected magnetic resonance (ODMR) spectra and developing a model that quantitatively describes its behavior.
Here, we measure the ODMR spectra of $\mathrm{V_B^-}$ in $\mathrm{h}^{10}\mathrm{B}^{15}\mathrm{N}$ over a wide magnetic field range, including the ground state level anti-crossing (GSLAC), and compare them with the results of the Lindblad-based simulation that considers a single electron spin and three neighboring $^{15}\mathrm{N}$ nuclear spins. 
Our simulation successfully reproduces the experimental spectra, including the vicinity of GSLAC. It can explain the overall behavior of the magnetic field dependence of the nuclear spin polarization estimated using the Lorentzian fitting of the spectra. Despite such qualitative agreement, we also demonstrate that the fitting methods cannot give accurate polarizations.
Finally, we discuss that symmetry-induced mechanisms of $\mathrm{V_B^-}$ limit the maximum polarization. 
Our study is an essential step toward a quantitative understanding of DNP using defects in hBN and its quantum applications.
\end{abstract}

\maketitle

\section{\label{sec:Intro}Introduction}
Spin defects in wide-bandgap semiconductors have attracted attention as a platform for developing quantum technologies~\cite{weber2010quantum,atature2018material,awschalom2018quantum,wolfowicz2021quantum}. 
For example, spin defects in diamond~\cite{childress2013diamond,rondin2014magnetometry,schirhagl2014nitrogen,casola2018probing} and SiC~\cite{castelletto2020silicon} are extensively studied for applications in quantum sensing, quantum simulation, and quantum communication due to their optical addressability and long coherence times from cryogenic to room temperatures. 
Recently, the boron vacancy ($\mathrm{V_B^-}$) in hexagonal boron nitride (hBN) was discovered to have electron spins that can be manipulated at room temperature~\cite{gottscholl2020initialization,gottscholl2021room}. 
Since hBN is a van der Waals (vdW) crystal, $\mathrm{V_B^-}$ has the following advantages in its quantum sensor application to materials:
(i) $\mathrm{V_B^-}$ can be brought within a few atomic layers of a measurement target through stamping methods~\cite{zhou2024sensing}. 
(ii) $\mathrm{V_B^-}$ can be created at arbitrary two-dimensional positions by irradiating hBN with a focused ion beam~\cite{sasaki2023magnetic,liang2023high}. 
(iii) hBN is thin, small, easy to process, and readily integrated into various devices. 
Currently, $\mathrm{V_B^-}$ is beginning to be used as a quantum sensor to explore the physical properties of materials~\cite{kumar2022magnetic,healey2023quantum,huang2022wide,zhou2024sensing}.

Both nitrogen and boron atoms have nuclear spins, which couple to the $\mathrm{V}_\mathrm{B}^-$ electron spin via hyperfine interactions. 
In general, nuclear spins in solids have longer coherence times than electron spins. 
In diamonds, for instance, they are frequently utilized as quantum memories~\cite{jiang2009repetitive,neumann2010single,waldherr2014quantum,taminiau2014universal,lovchinsky2016nuclear}. 
In such cases, the nuclear spins are manipulated and read out via electron spins.
Realizing such nuclear spin control and application with quantum defects in hBN~\cite{gao2022nuclear,gong2024isotope,ru2024robust} requires understanding the hyperfine interactions with two-dimensionally distributed nuclear spins and the phenomena arising from them.

Dynamic nuclear polarization (DNP) is a phenomenon driven by hyperfine interactions and is used for initializing and reading out nuclear spins.
Recently, several papers~\cite{gao2022nuclear,sasaki2023nitrogen,clua2023isotopic,gong2024isotope,ru2024robust} reported that the high polarization of $\mathrm{V}_\mathrm{B}^-$ electron spins induced by optical pumping is transferred to the three adjacent nitrogen nuclear spins. 
Gao \textit{et al.}~\cite{gao2022nuclear} demonstrated DNP and coherent control of nitrogen nuclear spins in hBN with natural isotope composition ratios (h$\mathrm{^{nat}B^{nat}N}$).
Ru \textit{et al.}~\cite{ru2024robust} reported DNP in h$\mathrm{^{nat}B^{nat}N}$ in the magnetic field of 25--100~mT and 140-- 200~mT, excluding the ground state level anti-crossing (GSLAC) of $\mathrm{V}_\mathrm{B}^-$. 
They proposed a model that considers a single electron spin, three nitrogen nuclear spins, and optical transitions and found that the calculated results of the Lindblad equation are qualitatively consistent with experimentally estimated polarization.
Some of the present authors~\cite{sasaki2023nitrogen}, Clua-Provost~\textit{et al.}~\cite{clua2023isotopic}, and Gong~\textit{et al.}~\cite{gong2024isotope} reported DNP in isotopically engineered hBN ($\mathrm{h^{nat}B^{15}N}$ or $\mathrm{h}^{10}\mathrm{B}^{15}\mathrm{N}$). 
The optically detected magnetic resonance (ODMR) spectra of $\mathrm{V}_\mathrm{B}^-$ in such isotopically engineered hBN are more straightforward than those of h$\mathrm{^{nat}B^{nat}N}$. 
Based on the simplified ODMR spectra, Clua-Provost~\textit{et al.}~\cite{clua2023isotopic} estimated the polarization of nitrogen nuclear spins in a wide range of magnetic fields (10--150~mT), including GSLAC, and compared the results with a simulation similar to that of Ru \textit{et al.}~\cite{ru2024robust}.

This paper first presents high-resolution and accurate ODMR spectra $\mathrm{V_B^-}$ in $\mathrm{h}^{10}\mathrm{B}^{15}\mathrm{N}$ obtained between 10 and 150~mT.
Second, we show that the estimated nuclear polarizations using the Lorentzian fitting of the ODMR spectra are qualitatively consistent with the true (genuine) polarization in the Lindblad-based simulation of a model that considers a single electron spin, three adjacent nitrogen nuclear spins, and optical transitions; that is, the simulation can explain the overall behavior of the polarization depending on the magnetic fields.
Third, we prove that the simulation considering microwave and inhomogeneous broadening reproduces the experimental spectra quantitatively over a wide range of magnetic fields, including GSLAC.
Nevertheless, we also demonstrate that the conventional Lorentzian fitting method is not sufficiently accurate to estimate the true polarization.
Finally, based on the simulation, we discuss that dynamics related to defect symmetry such as coexistence of flip-flop and flip-flip interactions and nuclear spin entanglement limit the maximum polarization.

This paper is organized as follows. 
Section \ref{sec:DNP_mechanism} introduces the fundamental properties of $\mathrm{V_B^-}$ in $\mathrm{h}^{10}\mathrm{B}^{15}\mathrm{N}$ and the hyperfine interactions with $^{15}\mathrm{N}$. Then, we explain the mechanism of DNP in $\mathrm{h}^{10}\mathrm{B}^{15}\mathrm{N}$. 
Section \ref{sec:Numerical_sim} describes our simulation based on the Lindblad equation. 
Section \ref{sec:Exp_setup} outlines our experimental setup and sample specifications. 
Section \ref{sec:Result_discuss} discusses the experimental and simulation results and their implications. 
Section \ref{sec:Conclusion} concludes the study.

\section{\label{sec:DNP_mechanism} DNP in hexagonal boron nitride}
\begin{figure}
    \centering
    \includegraphics[width=\linewidth]{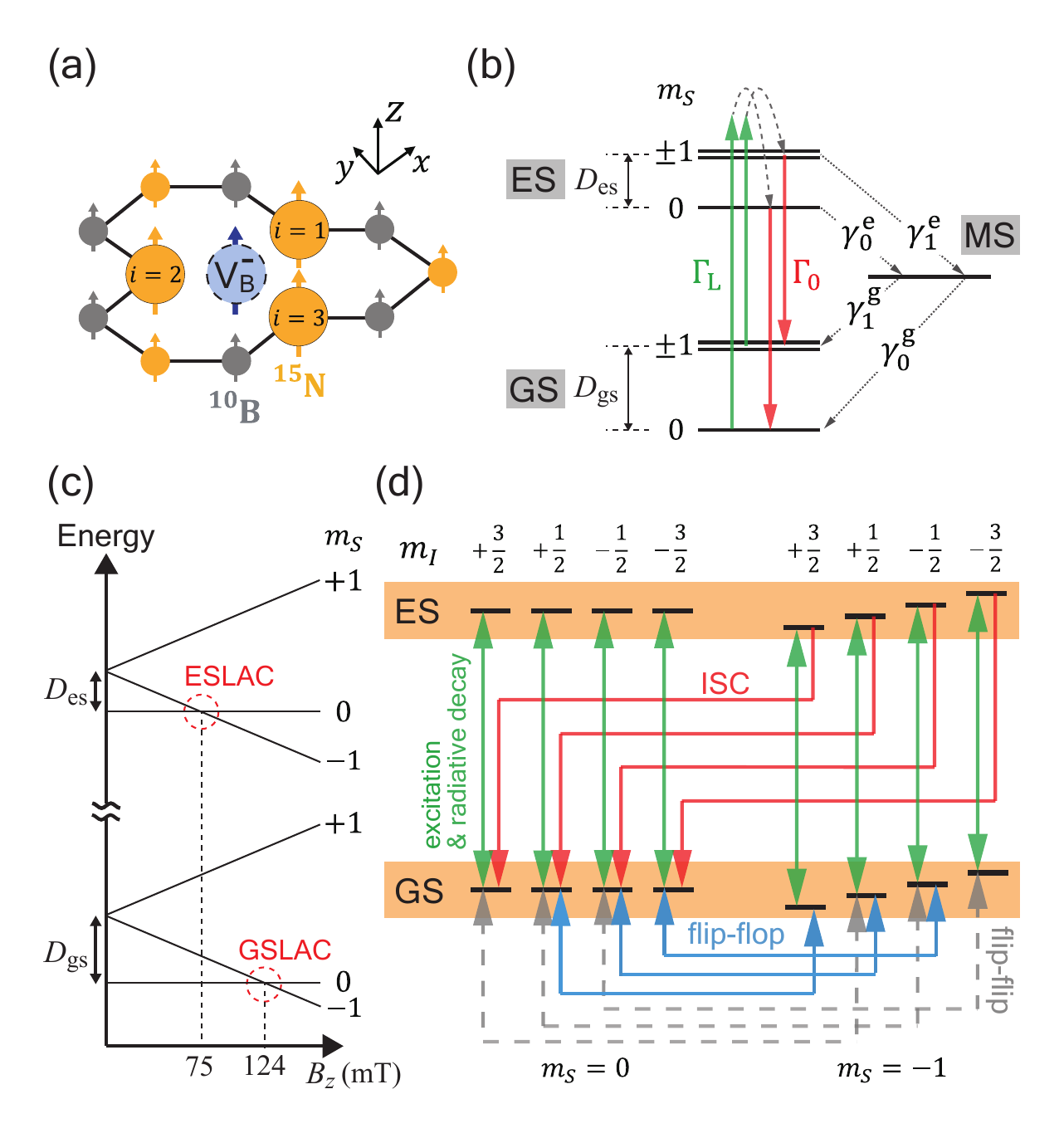}
    \caption{
    (a) Schematic of $\mathrm{V_{B}^{-}}$ in $\mathrm{h}^{10}\mathrm{B}^{15}\mathrm{N}$. 
    The $x$-axis is defined along the $a$-axis and the $z$-axis along the $c$-axis of the $\mathrm{h}^{10}\mathrm{B}^{15}\mathrm{N}$ crystal.
    (b) Energy level structure and optical transitions of $\mathrm{V_{B}^{-}}$. The relevant transition rates $\Gamma_\mathrm{L}$, $\Gamma_\mathrm{0}$, $\gamma_{0}^{\mathrm{e}}$, $\gamma_{1}^{\mathrm{e}}$, $\gamma_{0}^{\mathrm{g}}$, and $\gamma_{1}^{\mathrm{g}}$ are depicted.
    (c) Magnetic field dependence of $\mathrm{V_{B}^{-}}$ energy levels. 
    In the present experiment, the zero-field splitting is estimated as $D_{\mathrm{gs}} = 3.47$ GHz for the GS and $D_{\mathrm{es}} \sim 2.09$ GHz for the ES. 
    Level anti-crossings (LAC) between $m_S = 0$ and $m_S = -1$ occur at $B_z \sim 124$ mT (GS) and $B_z \sim 75$ mT (ES).
    (d) DNP mechanism. 
    Green arrows indicate optical excitation from GS to ES and radiative transitions from ES to GS. Red arrows represent ISC from ES $m_S = \pm 1$ to GS $m_S = 0$. 
    Light blue arrows correspond to flip-flop transitions $\ket{m_S = 0, m_I} \leftrightarrow \ket{m_S = -1, m_I + 1}$. 
    Gray dashed arrows indicate flip-flip transitions $\ket{m_S = 0, m_I + 1} \leftrightarrow \ket{m_S = -1, m_I}$.
    }
    \label{fig:intro}
\end{figure}

hBN is a vdW material composed of a hexagonal lattice arrangement of boron and nitrogen atoms, as shown in Fig.~\ref{fig:intro}(a). 
It is a wide-gap semiconductor with a bandgap of $\sim 6.0$ eV~\cite{zunger1976optical}. 
The negatively charged vacancy at the boron site is referred to as $\mathrm{V_{B}^-}$.
The electrons localized in this defect form multiple levels within the band gap, such as the orbital ground state (GS) and excited state (ES), which are spin triplets, and the metastable singlet state (MS) [Fig.~\ref{fig:intro}(b)].
When $\mathrm{V_{B}^-}$ is in the GS (ES) state, the electron spin levels are split by $D_{\mathrm{gs}} = 3.47$~GHz ($D_{\mathrm{es}} = 2.09$~GHz) in a zero magnetic field.
The electron spin has a quantized axis along the $z$-direction (the $c$-axis of the hBN crystal) [Fig.~\ref{fig:intro}(a)], and the magnetic quantum number $m_S$ is defined as the projection of the spin angular momentum onto the $z$-axis.
We can excite the transition from GS to ES in a spin-conserving manner ($\Delta m_S = 0$) by irradiating with a green laser (transition rate:~$\Gamma_\mathrm{L}$), as shown in Fig.~\ref{fig:intro}(b).
The relaxation from the ES to GS includes spin-conserving radiative relaxation, which involves a direct transition accompanied by the photoluminescence (PL) with a transition rate of ~$\Gamma_\mathrm{0}$, and spin-nonconserving non-radiative relaxation via intersystem crossing (ISC) through MS. 
The transition rates in this non-radiative paths are $\gamma_{0}^{\mathrm{e}}$, $\gamma_{1}^{\mathrm{e}}$, $\gamma_{0}^{\mathrm{g}}$, and $\gamma_{1}^{\mathrm{g}}$, as defined in Fig.~\ref{fig:intro}(b). 
Since $\gamma_{0}^{\mathrm{e}} < \gamma_{1}^{\mathrm{e}}$~\cite{baber2021excited,clua2024spin,lee2024intrinsic} and $\gamma_{0}^{\mathrm{g}} > \gamma_{1}^{\mathrm{g}}$~\cite{clua2024spin,lee2024intrinsic}, the non-radiative relaxation process allows us to polarize, or initialize, the electron spin state into $m_S=0$ through continuous optical excitation.
Furthermore, due to these relaxation processes, the PL intensity for the $m_S = \pm1$ states is weaker than that of the $m_S = 0$ state. 
Therefore, we can read out the electron spin state by referencing the difference in the PL intensity.

The electron spin of $\mathrm{V_{B}^-}$ in $\mathrm{h}^{10}\mathrm{B}^{15}\mathrm{N}$ is coupled to the three adjacent $\mathrm{^{15}N}$ nuclear spins via hyperfine interactions. 
In this paper, the three adjacent $\mathrm{^{15}N}$ atomic nuclei are indexed as $i = 1, 2$, and $3$, as shown in Fig. \ref{fig:intro}(a). 
The Hamiltonian considering the electron spin and the three nuclear spins is given by:
\begin{align}
&\hat{H}_{\mathrm{v}} 
= D_{\mathrm{v}}{\hat{S_z}}^2 + \gamma_{\mathrm{e}} B_z \hat{S}_z - \sum_{i=1}^{3} \gamma_{\mathrm{n}} B_z \hat{I}_z^{(i)} +  \sum_{i=1}^{3} \hat{\bm{S}} \bar{A}_{\mathrm{v}}^{(i)} \hat{\bm{I}}^{(i)}, \notag\\
&\sum_{i=1}^{3} \hat{\bm{S}} \bar{A}_{\mathrm{v}}^{(i)} \hat{\bm{I}}^{(i)} 
= \sum_{i=1}^{3} \sum_{j=x,y,z} \sum_{k=x,y,z} A_{\mathrm{v},jk}^{(i)} \hat{S}_j \hat{I}_k^{(i)},
\label{eq:Hamiltonian_wo_E_theta}
\end{align}
where $\mathrm{v} = \mathrm{gs~or~es}$ denotes the orbital levels, $B_z$ is the magnetic field along the $z$-axis, $\hat{\bm{S}}$ and $\hat{S}_z$ are the electron spin operators of $\mathrm{V_{B}^-}$ with $S = 1$, and $\hat{\bm{I}}^{(i)}$ and $\hat{I}_z^{(i)}$ are the nuclear spin operators of the nearest-neighbor $\mathrm{^{15}N}$ nuclei with $I^{(i)} = 1/2$. 
$\gamma_{\mathrm{e}} = 28$ MHz/mT is the gyromagnetic ratio of the electron spin, and $\gamma_{\mathrm{n}} = -4.3$ kHz/mT is the gyromagnetic ratio of the $\mathrm{^{15}N}$ nuclear spin.
The term $\sum_{i=1}^{3} \hat{\bm{S}} \bar{A}_{\mathrm{v}}^{(i)} \hat{\bm{I}}^{(i)}$ represents the hyperfine interactions between $\mathrm{V_{B}^-}$ and the nearest-neighbor $\mathrm{^{15}N}$ nuclei, where $A_{\mathrm{v},jk}^{(i)}$ ($j=x,y$, and $z$, $k=x,y$, and $z$) are the components of the hyperfine interaction tensor $\bar{A}_{\mathrm{v}}^{(i)}$.

In most of the magnetic fields used in this study, the condition $|D_{\mathrm{v}}  - \gamma_{\mathrm{e}} B_z| \gg |A_{\mathrm{v},jk}^{(i)}| \sim 100$~MHz is satisfied, allowing the secular approximation: $ \sum_{i=1}^{3} \hat{\bm{S}} \bar{A}_{\mathrm{v}}^{(i)} \hat{\bm{I}}^{(i)} \approx \sum_{i=1}^{3} A_{\mathrm{v},zz}^{(i)} \hat{S}_{z} \hat{I}_{z}^{(i)} = A_{\mathrm{v},zz} \hat{S}_{z} (\sum_{i=1}^{3} \hat{I}_{z}^{(i)})$.
Thus, the states $\ket{m_S, m_I = \sum_{i=1}^{3} {m}_{I}^{(i)}}$ form a good basis. As a result, the energy levels of $\ket{m_S = \pm 1}$ are split by $|A_{\mathrm{v},zz}|$ for each of the four total nuclear spin states $m_I = \{+3/2, +1/2, -1/2, -3/2\}$.

On the other hand, when the condition $|D_{\mathrm{v}} - \gamma_{\mathrm{e}} B_z| \sim |A_{\mathrm{v},jk}^{(i)}|$ is satisfied, the off-diagonal components of the Hamiltonian in Eq.~\eqref{eq:Hamiltonian_wo_E_theta} cannot be neglected. 
This condition is called the level anti-crossing (LAC) [red circles in Fig.~\ref{fig:intro}(c)].
The hyperfine interaction term in Eq.~\eqref{eq:Hamiltonian_wo_E_theta} can be rewritten as~\cite{gong2024isotope}:
\begin{multline}
\sum_{i=1}^{3} \hat{\bm{S}} \bar{A}_{\mathrm{v}}^{(i)} \hat{\bm{I}}^{(i)}
= \sum_{i=1}^{3} \{
A_{\mathrm{v},zz} \hat{S}_{z} \hat{I}_{z}^{(i)}
+ (A_{\mathrm{v},+}^{(i)} \hat{S}_{+} \hat{I}_{-}^{(i)} + h.c.) \\
+ (A_{\mathrm{v},-}^{(i)} \hat{S}_{+} \hat{I}_{+}^{(i)} + h.c.)
\},
\label{eq:HF_hc}
\end{multline}
where $\hat{S}_{\pm} \equiv \hat{S}_{x} \pm i \hat{S}_{y}$, $\hat{I}_{\pm}^{(i)} \equiv \hat{I}_{x}^{(i)} \pm i \hat{I}_{y}^{(i)}$, and
\begin{gather}
A_{\mathrm{v},+}^{(i)} = \dfrac{A_{\mathrm{v},xx}^{(i)} + A_{\mathrm{v},yy}^{(i)}}{4}, \notag\\
A_{\mathrm{v},-}^{(i)} = \dfrac{A_{\mathrm{v},xx}^{(i)} - A_{\mathrm{v},yy}^{(i)}}{4} + \dfrac{A_{\mathrm{v},xy}^{(i)}}{2i}.
\label{eq:def_Aplus_Aminus}
\end{gather}
In the derivation of Eqs.~\eqref{eq:HF_hc} and \eqref{eq:def_Aplus_Aminus}, the mirror symmetry of $\mathrm{V}_\mathrm{B}^-$ concerning the $x$-$y$ plane is considered, leading to $A_{\mathrm{v},xz}^{(i)} = A_{\mathrm{v},yz}^{(i)} = A_{\mathrm{v},zx}^{(i)} = A_{\mathrm{v},zy}^{(i)} = 0$~\cite{sasaki2023nitrogen,gong2024isotope}. The second term of the right-hand side of Eq.~\eqref{eq:HF_hc} represents flip-flop interactions and causes transitions between $\ket{m_S = 0, m_I} \leftrightarrow \ket{m_S = -1, m_I + 1}$ [light blue arrows in Fig.~\ref{fig:intro}(d)]. The third term represents flip-flip interactions, causing transitions between $\ket{m_S = 0, m_I + 1} \leftrightarrow \ket{m_S = -1, m_I}$ [gray dashed arrows in Fig.~\ref{fig:intro}(d)].

Considering the $D_{3h}$ symmetry of the hyperfine interactions in GS~\cite{gao2022nuclear}, the following simple expression can be obtained (see Appendix \ref{sec:Symmetry_hf} for details):
\begin{gather}
A_{\mathrm{gs},+}^{(i)} = \dfrac{A_{\mathrm{gs},xx} + A_{\mathrm{gs},yy}}{4}, \notag\\
A_{\mathrm{gs},-}^{(i)} = \dfrac{A_{\mathrm{gs},xx} - A_{\mathrm{gs},yy}}{4} e^{2i\phi^{(i)}}, \notag\\
\phi^{(1)} = 0, \phi^{(2)} = -2\pi/3, \phi^{(3)} = -\phi^{(2)},
\label{eq:hf_d3h}  
\end{gather}
where $A_{\mathrm{gs},xx} \equiv A_{\mathrm{gs},xx}^{(1)} <0$ and $A_{\mathrm{gs},yy} \equiv A_{\mathrm{gs},yy}^{(1)} <0$, leading to $|A_{\mathrm{gs},+}^{(i)}| > |A_{\mathrm{gs},-}^{(i)}|$.
A similar symmetry argument can be applied to ES, yielding $|A_{\mathrm{es},+}^{(i)}| > |A_{\mathrm{es},-}^{(i)}|$, as Appendix \ref{sec:Symmetry_hf} describes.
Therefore, the flip-flop interaction is dominant.

Under GSLAC or excited-state LAC (ESLAC) conditions, the flip-flop interaction induces the transition $\ket{m_S = 0, m_I} \leftrightarrow \ket{m_S = -1, m_I + 1}$. 
Optical excitation further drives transitions such as $\ket{m_S = -1, m_I + 1} \rightarrow \ket{m_S = 0, m_I + 1}$, initializing the electron spin while simultaneously experiencing the flip-flop interaction. 
This process increases the total angular momentum of the system, resulting in nuclear spin polarization toward increasing $m_I$.
In contrast, the flip-flip interaction induces the transition $\ket{m_S = 0, m_I+1} \leftrightarrow \ket{m_S = -1, m_I}$, leading to polarization toward decreasing $m_I$, causing depolarization.

The flip-flop and flip-flip transitions are maximized under the following conditions, respectively,
\begin{gather}
\textrm{(flip-flop:)} \;\;\;\; E_{\mathrm{v},0,m_I} = E_{\mathrm{v},-1,m_I+1},
\label{eq:LAC_condition_flipflop}  \\
\textrm{(flip-flip:)} \;\;\;\; E_{\mathrm{v},0,m_I+1} = E_{\mathrm{v},-1,m_I},
\label{eq:LAC_condition_flipflip}
\end{gather}
where $m_I = \{+1/2, -1/2, -3/2\}$, and $E_{\mathrm{v},m_S,m_I} \equiv \bra{m_S,m_I} \hat{H}_{\mathrm{v}} \ket{m_S,m_I}$ represents the diagonal elements of the Hamiltonian in Eq.~\eqref{eq:Hamiltonian_wo_E_theta}.
The magnetic fields that satisfy conditions Eqs.~\eqref{eq:LAC_condition_flipflop} and \eqref{eq:LAC_condition_flipflip} are as follows:
\begin{align}
\textrm{(flip-flop:)} \;\;\;\; B_{+-,\mathrm{v},m_I} &= \dfrac{D_{\mathrm{v}} + (m_I+1) A_{\mathrm{v},zz}}{\gamma_{e} + \gamma_{n}},
\label{eq:Bz_LAC_condition_flipflop}  \\
\textrm{(flip-flip:)} \;\;\;\; B_{++,\mathrm{v},m_I} &= \dfrac{D_{\mathrm{v}} + m_I A_{\mathrm{v},zz}}{\gamma_{e} + \gamma_{n}}.
\label{eq:Bz_LAC_condition_flipflip}
\end{align}
These equations tell that the magnetic fields that maximize the flip-flop and flip-flip transitions differ.
By using zero-field splitting and hyperfine interactions (listed in Table~\ref{tb:hyperfine_param}, introduced later), we can calculate the specific field conditions.
Then, we derive the range of the GSLAC by the upper and lower limits of these magnetic fields as follows:
\begin{gather}
B_\mathrm{GSLAC,low} < B_z < B_\mathrm{GSLAC,up}, \notag \\
B_\mathrm{GSLAC,low} \equiv B_{+-,\mathrm{gs},+1/2} - \dfrac{|A_{\mathrm{gs},yy}|}{\gamma_{e} + \gamma_{n}} = 115.8 \; \mathrm{mT}, \notag\\
B_\mathrm{GSLAC,up} \equiv B_{++,\mathrm{gs},-3/2} + \dfrac{|A_{\mathrm{gs},yy}|}{\gamma_{e} + \gamma_{n}} = 131.5 \; \mathrm{mT}.
\label{eq:def_Bz_GSLAC}
\end{gather}

\section{\label{sec:Numerical_sim}Simulation based on Lindblad equation}
One of the purposes of this study is to investigate how far the experimental results can be explained by the model based on the three $\mathrm{^{15}N}$ nuclear spins adjacent to the $\mathrm{V_{B}^-}$ in $\mathrm{h}^{10}\mathrm{B}^{15}\mathrm{N}$.
In this subsection, we introduce the Lindblad equation that considers the spin Hamiltonian and optical transitions described in the previous subsection and explain how to calculate the nuclear spin polarization and ODMR spectra.

The density matrix $\hat{\rho}$ in the present model can be expressed as follows:
\begin{gather}
\hat{\rho} = \sum_{m,n}\rho_{mn}\ket{m}\bra{n}, \notag\\
\ket{m}, \ket{n} = \ket{level, m_S, m_I^{(i)}},
\label{eq:def_density_matrix}
\end{gather}
where $\ket{level, m_S, m_I^{(i)}} \equiv \ket{level, m_S, m_I^{(1)}, m_I^{(2)}, m_I^{(3)}}$ ($i = 1$, 2, and $3$).
For $level = \mathrm{GS~or~ES}$, the basis states $\ket{level, m_S, m_I^{(i)}}$ are labeled by $m_S = \{+1,0,-1\}$ and $m_I^{(i)} = \{+1/2,-1/2\}$.
The basis states $\ket{\mathrm{MS}, m_S, m_I^{(i)}}$ are labeled by $m_S = \{0\}$ and $m_I^{(i)} = \{+1/2,-1/2\}$.
$\hat{\rho}$ is a $56 \times 56$ matrix ($56 = (2\times3+1\times1)\times2^3$), and it follows the Lindblad equation described below.
\begin{equation}
\dfrac{\partial \hat{\rho}}{\partial t} = -\dfrac{i}{\hbar} [\hat{H}, \hat{\rho}] + \hat{\mathscr{D}} (\hat{\rho}).
\label{eq:Lindblad}
\end{equation}
The Hamiltonian $\hat{H}$ in Eq.~\eqref{eq:Lindblad} is expressed as follows:
\begin{alignat}{2}
&\hat{H} & &= \hat{H}_{\mathrm{gs}} \oplus \hat{H}_{\mathrm{es}} \oplus \hat{H}_{\mathrm{ms}}, \notag\\
&\hat{H}_{\mathrm{(gs,es)}} & &= D_{\mathrm{(gs,es)}}{\hat{S_z}}^2 + \gamma_{\mathrm{e}} B (\hat{S}_z \cos{\theta} + \hat{S}_x \sin{\theta}) \notag\\
& & &+ E_{\perp} ({S_x}^2 - {S_y}^2) \notag\\
& & &- \sum_{i=1}^{3} \gamma_{\mathrm{n}}  B (\hat{I}_z^{(i)} \cos{\theta} + \hat{I}_x^{(i)} \sin{\theta}) + \hat{H}_{\mathrm{hf,(gs,es)}}, \notag\\
&\hat{H}_{\mathrm{hf,(gs,es)}} 
& &= \sum_{i=1}^{3} \{
A_{\mathrm{(gs,es)},zz} \hat{S}_{z} \hat{I}_{z}^{(i)}
+ A_\mathrm{(gs,es),+}^{(i)} (\hat{S}_{+} \hat{I}_{-}^{(i)} + h.c.) \notag\\
& & &+ A_\mathrm{(gs,es),-}^{(i)} (\hat{S}_{+} \hat{I}_{+}^{(i)} + h.c.)\}, \notag\\
&\hat{H}_{\mathrm{ms}} & &= -\sum_{i=1}^{3} \gamma_{\mathrm{n}}  B (\hat{I}_z^{(i)} \cos{\theta} + \hat{I}_x^{(i)} \sin{\theta})
\}.
\label{eq:Hamiltonian_full}
\end{alignat}
$\hat{H}_{\mathrm{gs}}$, $\hat{H}_{\mathrm{es}}$, and $\hat{H}_{\mathrm{ms}}$ are the spin Hamiltonians for GS, ES, and MS, respectively. 
$\hat{H}_{\mathrm{gs}}$ and $\hat{H}_{\mathrm{es}}$ are similar to Eq.~\eqref{eq:Hamiltonian_wo_E_theta}, but they additionally consider the tilt of the bias magnetic field $\theta$ and strain $E_\perp$.
$B$ denotes the magnitude of the magnetic field, and $B_z=B \cos\theta$.
The strain parameter $E_{\perp}=52$ MHz originates from lattice distortions~\cite{lyu2022strain} and the surrounding charge distribution~\cite{durand2023optically}.
$\hat{H}_{\mathrm{hf,(gs,es)}}$ represents the hyperfine interaction terms for GS and ES.
Unless otherwise specified, the simulation in this paper use the hyperfine parameters shown in Table~\ref{tb:hyperfine_param}. 
All hyperfine parameters, except for $A_{\mathrm{gs},zz}$, are based on the first-principles calculations~\cite{gao2022nuclear}, scaled by a factor of nitrogen isotope gyromagnetic ratios $\gamma_\mathrm{n}^\mathrm{^{15}N}/\gamma_\mathrm{n}^\mathrm{^{14}N}=-1.4$. 
$A_{\mathrm{gs},zz}=-64$ MHz is derived from the experimental data as described below.

\begin{table}[htbp]
\centering
\caption{Hyperfine parameters used in our simulation~\cite{gao2022nuclear}.}
\begin{tabular}{cc}
  Parameter &\;\; Value \\ \hline\hline
  $A_{\mathrm{gs},xx}$ & \;\; $-64$ MHz \\
  $A_{\mathrm{gs},yy}$ & \;\; $-125$ MHz \\
  $A_{\mathrm{gs},zz}$ & \;\; $-64$ MHz \\
  $A_{\mathrm{es},xx}^{(1)}$ & \;\; $-5.1$ MHz \\
  $A_{\mathrm{es},xx}^{(2,3)}$ & \;\; $-64$ MHz \\
  $A_{\mathrm{es},yy}^{(1)}$ & \;\; $-4.9$ MHz \\
  $A_{\mathrm{es},yy}^{(2,3)}$ & \;\; $-73$ MHz \\
  $A_{\mathrm{es},zz}^{(i=1,2,3)}$ & \;\; $-60$ MHz \\
  $A_{\mathrm{es},xy}^{(1)}$ & \;\; $0$ MHz \\
  $A_{\mathrm{es},xy}^{(2,3)}$ & \;\; $8.3$ MHz \\
  $A_{\mathrm{es},yx}^{(i=1,2,3)}$ & \;\; $-A_{\mathrm{es},xy}^{(i=1,2,3)}$ \\ \hline
  \label{tb:hyperfine_param}
\end{tabular}
\end{table}

\begin{table}[htbp]
\centering
\caption{Transition or relaxation rates used in our simulation. We neglect the electron spin and nuclear spin $T_1$ relaxations in ES.} %
\begin{tabular}{lrr}
  $k$ & Process\;\;\;\;\;\;\;\;\;\;\;\;\; &\; Rate $\Gamma_k$ \; \\ \hline\hline
  1 &\;\;Laser excitation: $\Gamma_{\mathrm{L}}$ & \;\;2.14 $\mu$s$^{-1}$ \\
  2 &\;\;Radiative decay: $\Gamma_{\mathrm{0}}$ & \;\;0.10 $\mu$s$^{-1}$ \\
  3 &\;\;Non-radiative decay: $\gamma_{0}^{\mathrm{e}}$ & \;\;2.0 ns$^{-1}$ \\
  4 &\;\;Non-radiative decay: $\gamma_{1}^{\mathrm{e}}$ & \;\;0.74 ns$^{-1}$ \\
  5 &\;\;Non-radiative decay: $\gamma_{0}^{\mathrm{g}}$ & \;\;38 $\mu$s$^{-1}$ \\
  6 &\;\;Non-radiative decay: $\gamma_{1}^{\mathrm{g}}$ & \;\;5.6 $\mu$s$^{-1}$ \\
  7,~8 &\;\; electron spin $T_1$ in GS: $1/3T_{1,\mathrm{gs}}$ & \;\; 1/(12 $\mathrm{\mu}$s)/3 \\
  9 &\;\;electron spin $T_2$ in GS: $1/T_{2,\mathrm{gs}}$ & \;\; 1/(180 ns)\\
  10 &\;\; electron spin $T_2$ in ES: $1/T_{2,\mathrm{es}}$ & \;\; 1/(2 ns)\\
  11,~12 &\;\; nuclear spin $T_1$ in GS: $1/2 T_{1,\mathrm{n}}$ & \;\; 1/(2 ms)/2\\
  13 &\;\; nuclear spin $T_2$ in GS and ES: $1/T_{2,\mathrm{n}}$ & \;\; 1/(200 $\mathrm{\mu}$s)\\ \hline
  \label{tb:decay_param}
\end{tabular}
\end{table}

The second term on the right-hand side of Eq. \eqref{eq:Lindblad} is expressed as follows:
\begin{equation}
\hat{\mathscr{D}} (\hat{\rho}) =
 \sum_{k} \Gamma_{k} \left( \hat{L}_k \hat{\rho} {\hat{L}_k}^{\dagger} - \dfrac{1}{2} \{{\hat{L}_k}^{\dagger}\hat{L}_k, \hat{\rho}\} \right).
\label{eq:L_rho}
\end{equation}
Here, $\hat{L}_k$ are jump operators describing non-Hermitian processes, and $\Gamma_k$ are the transition or relaxation rates. 
The optical transitions from $\ket{m}$ to $\ket{n}$ are included in Eq.~\eqref{eq:Lindblad} with a jump operator of $\hat{L}_k = \ket{n}\bra{m}$.
$T_1$ relaxations are included in Eq.~\eqref{eq:Lindblad} with a jump operator of $\hat{L}_k = \hat{S_{+}}/\sqrt{2}$, $\hat{S_{-}}/\sqrt{2}$, $\hat{I_{+}}$, or $\hat{I_{-}}$.
The phase relaxation in the superposition of $\ket{m}$ and $\ket{n}$ (e.g., $T_2$ relaxation) is included in Eq.~\eqref{eq:Lindblad} with a jump operator of $\hat{L}_k = \ket{m}\bra{m} - \ket{n}\bra{n}$.
As shown in Table~\ref{tb:decay_param}, we consider a total of 13 non-Hermitian processes ($k = 1,2,\cdots$, and $13$) in Eq.~\eqref{eq:Lindblad}. 
The processes corresponding to $k = 1,2,\cdots$, and $6$ represent the optical transitions depicted in Fig.~\ref{fig:intro}(b). 
The processes for $k = 7,8,\cdots$, and $13$ describe the $T_1$ and $T_2$ relaxations of the $\mathrm{V_{B}^-}$ electron spin and $\mathrm{^{15}N}$ nuclear spins.

Unless otherwise noted, we perform the simulation by inserting the parameters listed in Table~\ref{tb:decay_param} into Eq.~\eqref{eq:L_rho}.
All optical-transition parameters ($k = 1,2,\cdots$, and $6$) are determined by fitting the time-resolved PL (see Appendix~\ref{sec:trPL}).
The $T_1$ relaxation rates of the $\mathrm{V_{B}^-}$ electron spin in the GS ($k=7$ and 8) are experimentally determined, and the $T_2$ relaxation rate in the ES ($k=10$) is adopted from Ref.~\cite{clua2023isotopic}. 
The other relaxation rates ($k=9$, 11, 12, and $13$) are adjusted to match the simulation results with the experimental results.

Under the above conditions, we solve the steady-state solution of the Lindblad equation for the following two cases:
\begin{itemize}
\setlength{\parskip}{0cm}
\setlength{\itemsep}{0cm}
  \item[(A)]  Without microwave irradiation.
  \item[(B)] With microwave irradiation.
\end{itemize}
We now explain the two cases of (A) and (B) in turn.

For case (A), we numerically obtain the steady-state $\hat{\rho}_\mathrm{sat}$ with the Lindblad equation by converting Eq.~\eqref{eq:Lindblad} into a Liouville space form [see Appendix~\ref{sec:Lindblad_Liouville}]. 
We can obtain the polarization of the three adjacent $\mathrm{^{15}N}$ nuclear spins using the following formula:
\begin{gather}
P_\mathrm{sim,true} = \dfrac{\sum_{m_I} m_I \rho_{\mathrm{gs},m_I}}{(3/2) \sum_{m_I} \rho_{\mathrm{gs}, m_I}}, \notag\\
\rho_{\mathrm{gs},m_I} = \sum_{\sum m_I^{(i)} = m_I} \sum_{m_S} \bra{\mathrm{GS}, m_S, m_I^{(i)}} \hat{\rho}_\mathrm{sat}
\ket{\mathrm{GS}, m_S, m_I^{(i)}},
\label{eq:def_Pol_sim}
\end{gather}
where $P_\mathrm{sim,true}$ is a true (genuine) nuclear spin polarization in the simulation.
The present approach is similar to that proposed previously~\cite{ru2024robust,clua2023isotopic}.

In experiments, the polarization is usually estimated using the fitting of ODMR spectra with the sum of four Lorentzians~\cite{ru2024robust,sasaki2023nitrogen}.
We refer to this method as ``4-dip fitting'' and  discuss experimental polarization values $P_\mathrm{exp,4dip}$ and $P_\mathrm{exp,4dip,area}$ defined in Eqs.~\eqref{eq:def_Pol_exp} and \eqref{eq:def_Pol_area}, respectively, later.
This estimation assumes that the contrast or area of a decomposed spectral component is proportional to the nuclear spin population.
However, as we will discuss in detail in Sec.~\ref{sec:Comparison-based-on-the-4-dip-fitting}, consideration should be given that the estimation accuracy is degraded if the ODMR spectra are distorted by factors other than the nuclear spin population.

Now, we calculate the ODMR spectra in case (B) and compare them either directly or with the results of 4-dip fitting to make a fairer comparison with the experimental results.
We consider microwave irradiation by rewriting Eq.~\eqref{eq:Hamiltonian_full} as follows:
\begin{align}
\hat{H}_{\mathrm{(gs,es)}} &\rightarrow \hat{H}_{\mathrm{(gs,es)}}
+ B_{\mathrm{mw}} \cos{(2\pi f_\mathrm{mw}t)} (\gamma_{\mathrm{e}} \hat{S}_x + \gamma_{\mathrm{n}} \sum_{i=1}^{3} \hat{I}_x^{(i)}), \notag\\
\hat{H}_{\mathrm{ms}} &\rightarrow \hat{H}_{\mathrm{ms}} 
+ \sum_{i=1}^{3} \gamma_{\mathrm{n}} B_{\mathrm{mw}} \cos{(2\pi f_\mathrm{mw}t)} \hat{I}_x^{(i)},
\label{eq:add_MW}
\end{align}
where $B_{\mathrm{mw}}$ and $f_{\mathrm{mw}}$ are the microwave amplitude and frequency, respectively.
Using the $\hat{\rho}_\mathrm{sat}$ obtained in case (A) as the initial condition, we calculate the time evolution under microwave irradiation using Eq.~\eqref{eq:Lindblad}.
We numerically obtain the state at $t = 1$ $\mathrm{\mu s}$ when the system has sufficiently relaxed and regard this as the steady state under microwave irradiation.
The PL intensity $I_\mathrm{sim}$ under microwave irradiation is expressed as follows:
\begin{gather}
I_\mathrm{sim} = \Gamma_{0} \rho_\mathrm{es}, \notag\\
\rho_{\mathrm{es}} = \sum_{m_S, m_I^{(i)}} \bra{\mathrm{ES}, m_S, m_I^{(i)}} \hat{\rho}(t=\, 1 \, \mu s) \ket{\mathrm{ES}, m_S, m_I^{(i)}}. 
\label{eq:def_PL_sim}
\end{gather}
By calculating the PL intensity for each microwave frequency, we obtain  ODMR spectra.
By applying 4-dip fitting to these simulated spectra to emulate experimental polarization estimation, we obtain $P_\mathrm{sim,4dip}$.
In Sec.~\ref{sec:Comparison-based-on-the-4-dip-fitting}, we evaluate the accuracy of the 4-dip fitting  by how well $P_\mathrm{sim,4dip}$ agrees with $P_\mathrm{sim,true}$ [Eq.~\eqref{eq:def_Pol_sim}].

We performed numerical simulation on a workstation equipped with dual intel CPUs (56 cores in total).
A reasonable calculation speed was obtained using the Julia packages DifferentialEquations.jl and Distributed.jl to calculate time evolution.

\section{\label{sec:Exp_setup}Experimental method}
\begin{figure*}
    \centering
    \includegraphics[width=\linewidth]{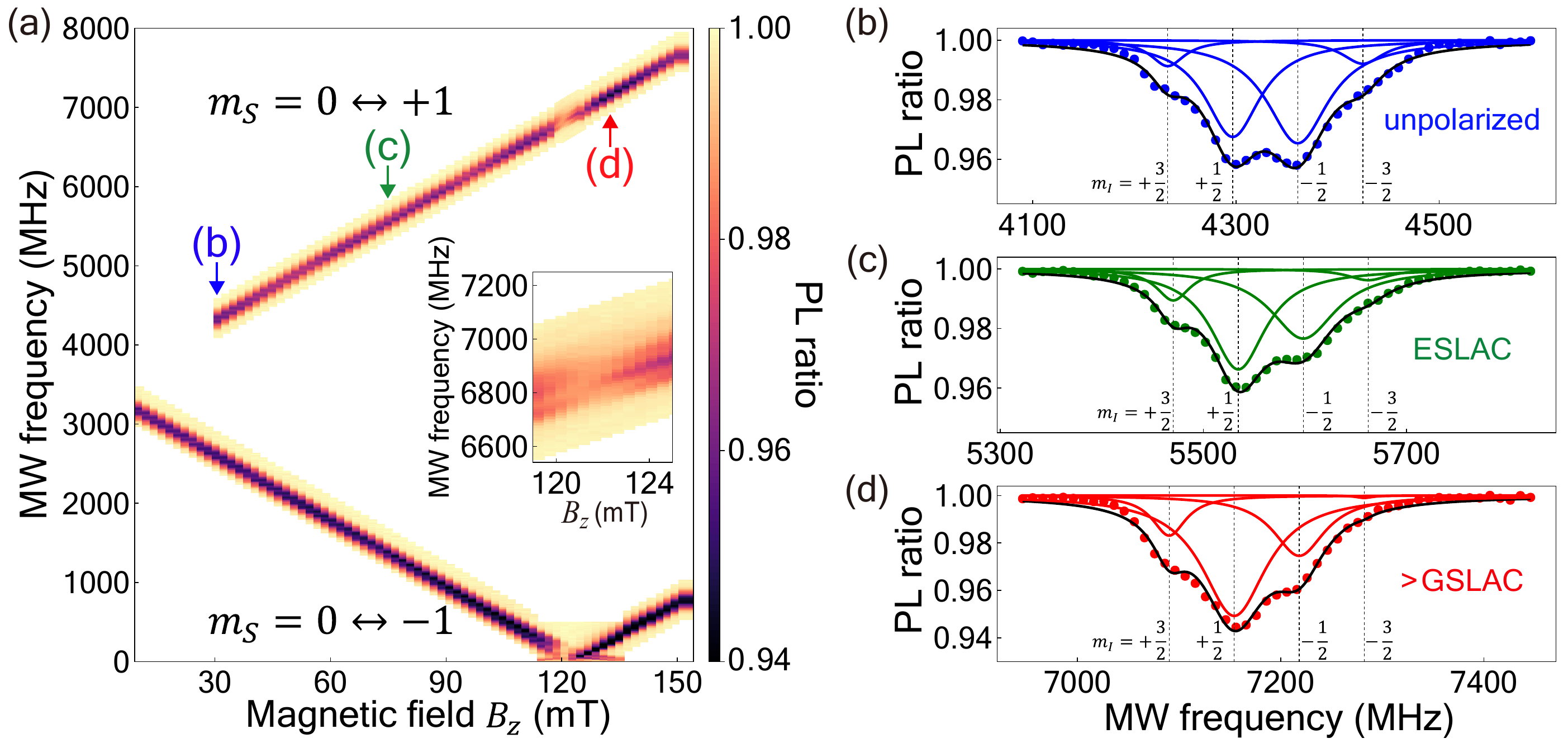}
    \caption{
    ODMR spectra at an optical power of 0.73~mW.
    (a) Magnetic field dependence of the spectra is shown on a color map. 
    The vertical axis shows the microwave frequency, the horizontal axis indicates the magnetic field $B_z$, and the color bar represents the PL ratio, which is the ratio of the PL intensity with and without microwaves. 
    The map shows magnetic resonances for both $m_S = 0 \leftrightarrow +1$ and $m_S = 0 \leftrightarrow -1$.
    (b)(c)(d) ODMR spectra corresponding to the $m_S = 0 \leftrightarrow +1$ transition at (b) $B_z = 30.7$~mT, (c) $B_z = 75.0$~mT, and (d) $B_z = 132.8$~mT.
    The vertical axis shows the PL ratio, and the horizontal axis shows the microwave frequency.
    The black solid lines represent the fitting results of four Lorentzians (``4-dip fitting''). 
    The decomposed Lorentzians are shown as solid blue, green, or red lines in (b), (c), and (d), respectively.
    }
    \label{fig:show_odmr}
\end{figure*}
We measure the ODMR of $\mathrm{V_{B}^-}$ in $\mathrm{h}^{10}\mathrm{B}^{15}\mathrm{N}$ synthesized using a metathesis reaction~\cite{sasaki2023nitrogen}.
The hBN flake, approximately 70~nm thick, was stamped onto an 800~nm wide gold wire on a sapphire substrate, which was fabricated using electron beam lithography (Elionix, ELS-F125).
We used a helium ion microscope (Carl Zeiss Microscopy, Orion Plus HIM) to irradiate a $(300~\mathrm{nm})^2$-sized spot with helium ions (acceleration voltage of 30~keV, irradiation dose of $10^{15}$~$\mathrm{/cm^2}$) to create a $\mathrm{V_{B}^-}$ ensemble~\cite{sasaki2023magnetic,gu2024systematic}.

Using a home-built confocal microscope~\cite{misonou2020construction} equipped with optical filters optimized for the PL wavelength of $\mathrm{V_{B}^-}$~\cite{gu2023multi}, we illuminated the $\mathrm{V_{B}^-}$ spot with a green laser (wavelength of 532~nm). 
The PL in the wavelength range from 750 to 1000 nm was detected using a single-photon counting module (Excelitas Technologies, SPCM-AQRH-16-FC-ND). 
Microwaves for manipulating the $\mathrm{V_{B}^-}$ electron spin were supplied to the gold wire via coaxial cables and coplanar waveguides from an analog signal generator (Keysight, E8257D). 
The microwave magnetic field was spatially concentrated around the thin gold wire ($\sim800$~nm), allowing us to drive the $\mathrm{V_{B}^-}$ electron spin without using a microwave amplifier. 
This configuration provides uniform microwave field strength over a wide frequency range from 1~MHz to 8~GHz.
A magnetic field parallel to the quantization axis of the $\mathrm{V_{B}^-}$ was applied in the range from 10 to 150~mT using an electromagnet (GMW, 5203). 
The alignment of the magnetic field was performed by moving a manual stage carrying the electromagnet.

\section{\label{sec:Result_discuss}Results and discussion}
\subsection{\label{subsec:ODMR_vs_Bz}Magnetic field dependence of ODMR spectra}
We measured the ODMR spectra for $m_S = 0 \leftrightarrow +1$ and $m_S = 0 \leftrightarrow -1$ at $B_z = 10$--$150$ mT. 
We present the result obtained at the optical power of 0.73~mW as functions of the microwave frequency and $B_z$ as a color map in Fig.~\ref{fig:show_odmr}(a).
The transitions between $m_S = 0 \leftrightarrow \pm1$ decrease the PL ratio, which is the ratio of the PL intensity with and without microwaves.
The upper (lower) branch corresponds to the resonance of $m_S = 0 \leftrightarrow +1$ ($m_S = 0 \leftrightarrow -1$). 
Each resonance frequency shift is equivalent to the product of the magnetic field and the gyromagnetic ratio $\gamma_{\mathrm{e}}$ due to the Zeeman splitting.
Note that the ODMR signal of the ES is small enough to be ignored at the microwave and optical power in this experiment ($<0.2$\% at $B_z = 0$~mT).

We focus on the ODMR spectra at several characteristic magnetic fields, as indicated by the arrows (b), (c), and (d) in Fig.~\ref{fig:show_odmr}(a).
Figure~\ref{fig:show_odmr}(b) shows the spectrum at $B_z = 30.7$~mT, away from the LACs. 
The spectrum consists of four dips. 
They correspond to the electron spin resonances under $m_I = +3/2, +1/2, -1/2$, and $-3/2$.
The separation between the dips equals the hyperfine interaction strength of $|A_{\mathrm{gs},zz}|=64$~MHz, which is consistent with the previous work~\cite{sasaki2023nitrogen,clua2023isotopic,gong2024isotope}.
The symmetric shape of the spectrum tells that the nitrogen nuclear spins are not polarized.

\begin{figure*}[t]
    \centering
    \includegraphics[width=\linewidth]{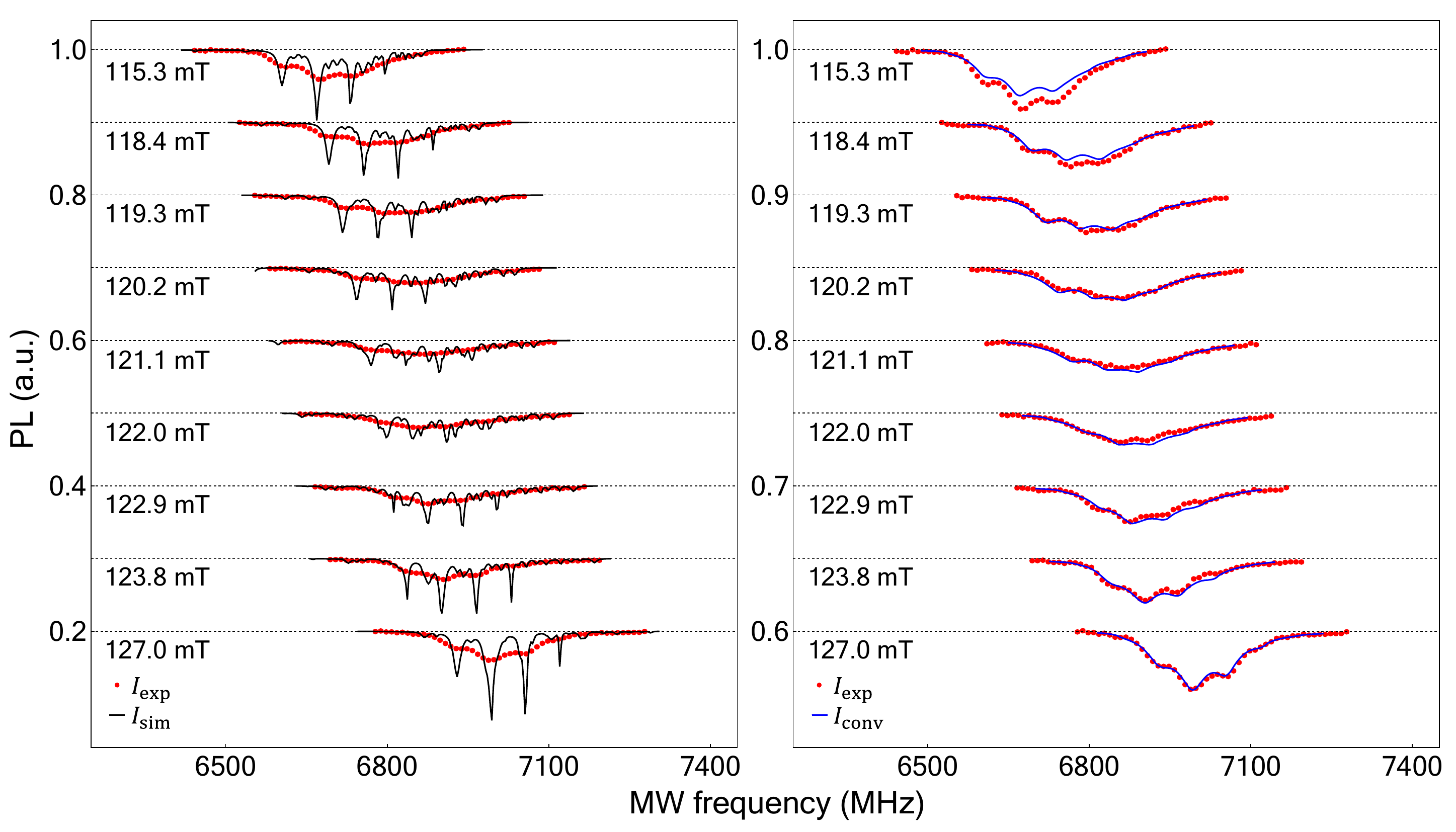}
    \caption{
    ODMR spectra near GSLAC at various magnetic fields $B_z$ between 115.3 to 127.0~mT under the optical power of 0.18~mW. 
    (Left panel) The experimental spectra $I_\mathrm{exp}$ (red markers) and the simulated spectra $I_\mathrm{sim}$ (black solid lines). 
    (Right panel) The red markers show $I_\mathrm{exp}$ (the data are the same as those in the left panel), and the simulated spectra considering inhomogeneous broadening $I_\mathrm{conv}$ are demonstrated by blue solid lines.
    In this simulation, we use $\theta = 0.6^{\circ}$, $\Gamma_\mathrm{L} = 2.14$~MHz, and $B_\mathrm{mw} = 0.25$~mT. 
    }
    \label{fig:odmr_GSLAC}
\end{figure*}

On the other hand, the ODMR spectra are not symmetric at specific magnetic fields.
Figure~\ref{fig:show_odmr}(c) displays the spectrum at $B_z = 75.0$~mT on the ESLAC; The dips for $m_I>0$ are enhanced, suggesting that the nuclear spins are polarized in the $m_I>0$ direction. 
Figure~\ref{fig:show_odmr}(d) shows the spectrum at $B_z = 132.8$~mT, slightly above the range of the GSLAC ($B_z > B_\mathrm{GSLAC,up}=131.5$~mT) [see Eq.~\eqref{eq:def_Bz_GSLAC}]. 
Again, as in Fig.~\ref{fig:show_odmr}(c), the asymmetric spectrum suggests the nuclear spin polarization.
Note that the spectra under the GSLAC [inset in Fig.~\ref{fig:show_odmr}(a)] are more complex and varied, making it difficult to estimate the polarization straightforwardly.
We will discuss this point in more detail later in Fig.~\ref{fig:odmr_GSLAC}.

To estimate the nuclear spin polarization, we use the 4-dip fitting.
The fitting by four Lorentzians (solid lines) reproduces the experimental ODMR spectra (markers) well, as presented in Figs.~\ref{fig:show_odmr}(b), (c), and (d).
The fitting yields the contrast $C_{m_I}$ and line width $\nu_{m_I}$ of each dip.
The area of each dip $A_{m_I}$ can be estimated from the contrast and line width.
In the conventional method, the nuclear spin polarization is calculated by considering that the area of each dip corresponds to the occupancy of each $m_I$.
However, the correspondence between the dip area and the occupancy does not always hold because the line width varies in a complex manner due to competition between microwave driving and other inhomogeneous broadening.
Therefore, in this study, we simply estimate nuclear spin polarization by assuming that the contrast of each dip corresponds to the occupancy of each $m_I$ using the following equation:
\begin{equation}
P_\mathrm{exp,4dip} = \dfrac{\sum_{m_I} m_I C_{m_I}}{(3/2) \sum_{m_I} C_{m_I}}.
\label{eq:def_Pol_exp}
\end{equation}
We obtain $P_\mathrm{exp,4dip} = 0.1\pm 0.1\%$, $14\pm1\%$, and $26\pm3\%$ for Figs.~\ref{fig:show_odmr}(b), (c), and (d), respectively.
The accuracy of the estimation method based on the 4-dip fitting, including the area-based one, is discussed later in Sec.~\ref{sec:Comparison-based-on-the-4-dip-fitting}.

\subsection{\label{subsec:Pol_vs_Bz}Magnetic field dependence of polarization}
\begin{figure}
    \centering
    \includegraphics[width=\linewidth]{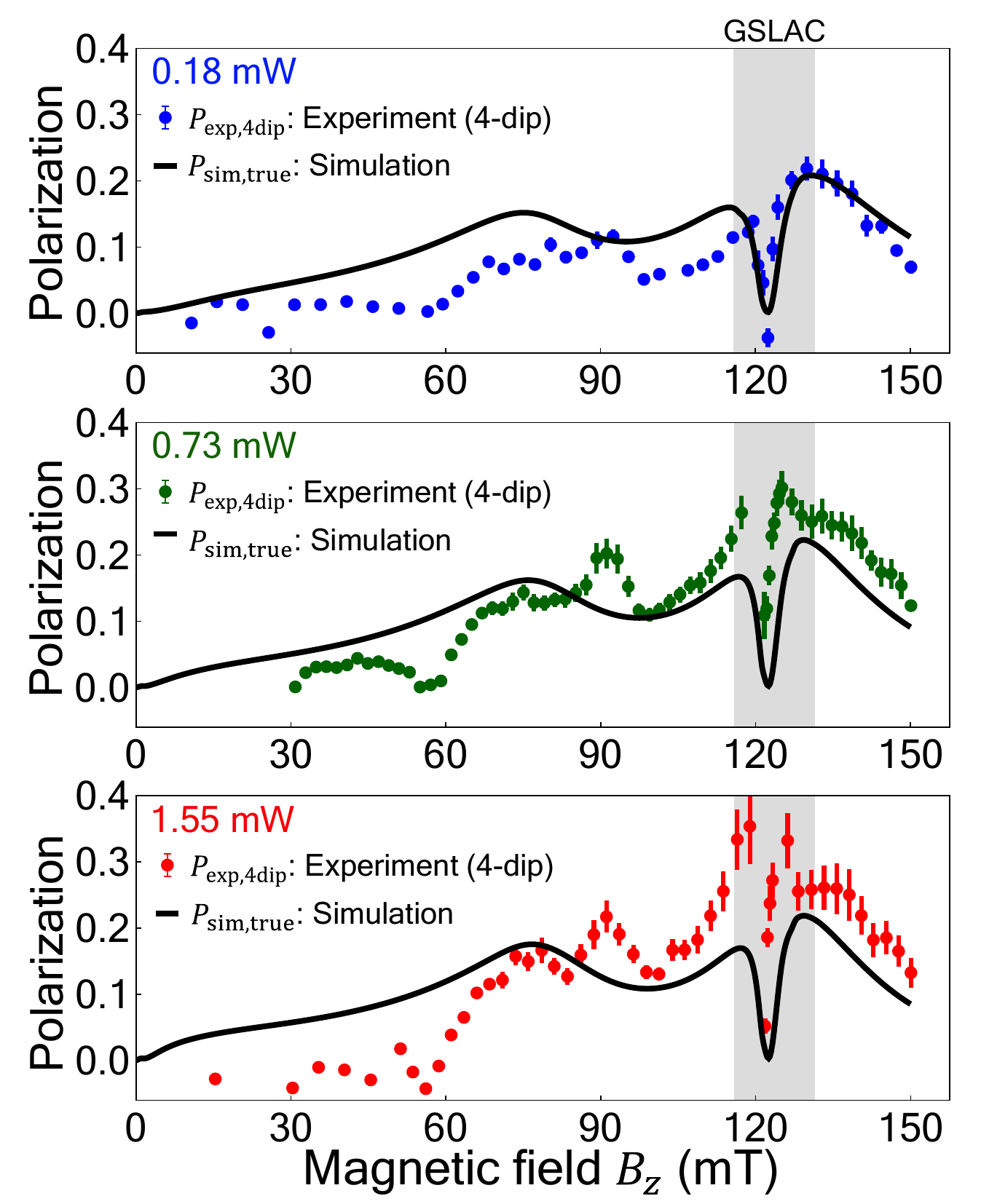}
    \caption{
    Magnetic field dependence of the polarization of the adjacent $^{15}\mathrm{N}$ nuclear spins.
    The top, middle, and bottom panels are obtained at optical powers of $0.18$, 0.73, and $1.55$ mW, respectively.
    Markers indicate the polarization $P_\mathrm{exp,4dip}$ estimated using the 4-dip fitting of ODMR spectra for $m_S = 0 \leftrightarrow +1$ and Eq.~\eqref{eq:def_Pol_exp}.
    The black solid lines are the true nuclear polarization $P_\mathrm{sim,true}$ [Eq.~\eqref{eq:def_Pol_sim}] in the steady state under continuous optical excitation in our simulation.
    The gray area indicates the range of GSLAC ($B_\mathrm{GSLAC,low} < B_z < B_\mathrm{GSLAC,up}$) [Eq.~\eqref{eq:def_Bz_GSLAC}]. 
    We use $\theta = 0.6^{\circ}$ in the simulation. 
    }
    \label{fig:pol_dep_Bz_dep_Laser}
\end{figure}

\begin{figure}
    \centering
    \includegraphics[width=\linewidth]{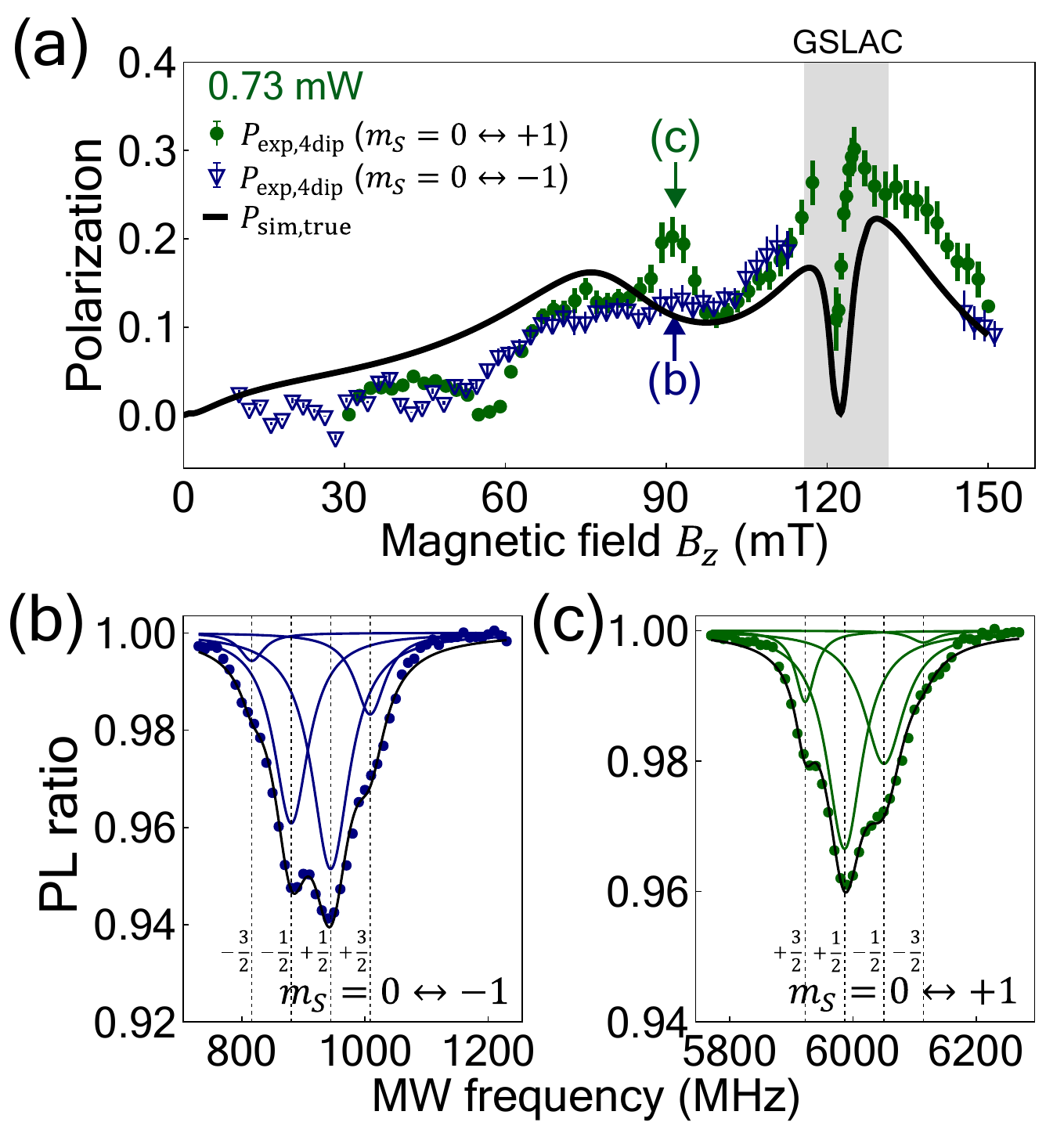}
    \caption{
    (a) Comparison of $P_\mathrm{exp,4dip}$ as a function of $B_z$ between $m_S = 0 \leftrightarrow +1$ (green) and $m_S = 0 \leftrightarrow -1$ (navy). 
    The optical power is fixed at 0.73~mW. 
    The black solid line presents the corresponding $P_\mathrm{sim,true}$.
    $P_\mathrm{exp,4dip}$ ($m_S = 0 \leftrightarrow +1$) and $P_\mathrm{sim,true}$ are the same as those shown in the middle panel of Fig.~\ref{fig:pol_dep_Bz_dep_Laser}.
    (b) ODMR spectrum and the result of the 4-dip fitting for $m_S = 0 \leftrightarrow -1$ (navy) at $B_z = 91.0$~mT. 
    (c) ODMR spectrum and the result of the 4-dip fitting for $m_S = 0 \leftrightarrow +1$ (green) at $B_z = 91.2$~mT.
    }
    \label{fig:pol_dep_Bz_dep_mS_pm1}
\end{figure}

We discuss the magnetic field dependence of the polarization of the adjacent $^{15}\mathrm{N}$ nuclear spins, estimated by analyzing the experimental ODMR spectra.
First, we focus on the result obtained at the optical power of 0.73~mW shown in the middle panel of Fig.~\ref{fig:pol_dep_Bz_dep_Laser}. 
The markers present the magnetic field dependence of the polarization $P_\mathrm{exp,4dip}$ [Eq.~\eqref{eq:def_Pol_exp}] obtained for the ODMR spectra shown in Fig.~\ref{fig:show_odmr}(a) using the 4-dip fitting.
$P_\mathrm{exp,4dip}$ increases around ESLAC ($\sim75$~mT) and GSLAC ($\sim120$~mT) except for the middle of GSLAC (gray area).

We can successfully reproduce a similar behavior using the simulation discussed in Sec.~\ref{sec:Numerical_sim}; The magnetic field dependence of $P_\mathrm{sim,true}$ [Eq.~\eqref{eq:def_Pol_sim}] is superposed by the black solid line in the same panel.
$P_\mathrm{exp,4dip}$ and $P_\mathrm{sim,true}$ behave similarly in the magnetic fields.
It is evident that the flip-flop interaction contributes to the increase in polarization.
Furthermore, we obtain a steep decrease in polarization $P_\mathrm{sim,true}$ at the middle of GSLAC ($B_z = 122.0$~mT). 
The flip-flip interaction for the ground state is most pronounced at this magnetic field.

Next, we show the optical power dependence.
The top and bottom panels of Fig.~\ref{fig:pol_dep_Bz_dep_Laser} are the results obtained at the optical powers of 0.18~mW and 1.55~mW, respectively.
Similar to the behavior in the middle panel (0.73~mW), the polarization $P_\mathrm{exp,4dip}$ changes around ESLAC and GSLAC in both cases.
Also, Fig.~\ref{fig:pol_dep_Bz_dep_Laser} indicates that the higher the optical power, the larger the polarization change appears near the ESLAC.

We remark that the 4-dip fitting analysis is not straightforward near GSLAC, shown in gray in Fig.~\ref{fig:pol_dep_Bz_dep_Laser}, because mixing electron and nuclear spins results in complex ODMR spectra.
Although our simulation depicted by the solid black lines qualitatively reproduces the trend under GSLAC, even this observation is theoretically non-trivial.
A more quantitative discussion and analysis will be made later in Secs.~\ref{subsec:odmr_at_gslac} and ~\ref{sec:Comparison-based-on-the-4-dip-fitting}.

So far, our simulation reproduces the observed behavior qualitatively well.  
At the end of this subsection, we point out an experimental result that cannot be reproduced.
Figure~\ref{fig:pol_dep_Bz_dep_mS_pm1}(a) shows the estimated polarization $P_\mathrm{exp,4dip}$ for $m_S = 0 \leftrightarrow +1$ and $m_S = 0 \leftrightarrow -1$ under a laser power of 0.73~mW. 
The polarizations obtained in these two transitions agree very well, except near $B_z \sim 91$~mT; A prominent peak appears at $B_z \sim 91$ mT only for the $m_S = 0 \leftrightarrow +1$ transition.
Such a behavior does not exist in the simulated $P_\mathrm{sim,4dip}$ shown by a solid line.
To examine this phenomenon more, we present the ODMR spectra for $m_S = 0 \leftrightarrow -1$ and $m_S = 0 \leftrightarrow +1$ at $B_z \sim 91$~mT in Figs.~\ref{fig:pol_dep_Bz_dep_mS_pm1}(b) and (c), respectively.
The ODMR contrasts corresponding to each nuclear spin state clearly differ between the two spectra.
We believe this is due to something we did not consider in our simulation model, such as nuclear spins other than the nearest-neighbor nitrogen atoms~\cite{haykal2022decoherence} or fast and complex optical transitions~\cite{ivady2020ab,reimers2020photoluminescence}.
We leave the identification of this cause as a future work.

\subsection{\label{subsec:odmr_at_gslac}ODMR spectra at GSLAC}
We discuss the ODMR spectra at GSLAC in detail and investigate their consistency with our simulation model.

First, we show the experimental ODMR spectra $I_\mathrm{exp}$ (red markers) and simulated spectra $I_\mathrm{sim}$ (black solid lines) at various magnetic fields $B_z$ between 115.3 and 127.0~mT in the left panel of Fig.~\ref{fig:odmr_GSLAC}.
As the magnetic field approaches $121$~mT, the experimental spectra become broad so that the four dips are almost indistinguishable.
Correspondingly, the simulation results indicate that the spectra shape becomes complex, and many dips appear around $121$~mT.

The above behavior is due to mixing the electron and nuclear spins in $\mathrm{V_{B}^-}$ at GSLAC.
The number of dips increases because the complex mixing increases the number of transitions between eigenstates.
Theoretically, strong mixing can occur under conditions where Eq.~\eqref{eq:def_Bz_GSLAC} is satisfied.
In the present case, the magnetic field conditions with strong mixing are estimated to be $B_z = 120.2$, $121.1$, and $122.0$~mT, which are consistent with experiments and simulations.
Such mixing is also observed in the LAC of diamond nitrogen-vacancy centers~\cite{auzinsh2019hyperfine,busaite2020dynamic}.

Next, we discuss the differences between the experimental and simulated spectra.
The simulated spectra are sharper than the experimental ones due to hyperfine interactions not fully considered in our simulations.
Our model described in Sec.~\ref{sec:Numerical_sim} considers line broadening due to microwave or laser irradiation~\cite{dreau2011avoiding}, $T_2$ relaxation, and $T_1$ relaxation (see Table \ref{tb:decay_param}).
On the other hand, it does not count inhomogeneous linewidths due to hyperfine interactions with boron nuclear spins, which is the main reason for the experimental linewidth broadening of $\mathrm{V_{B}^-}$~\cite{haykal2022decoherence}.
Accurately accounting for multiple boron spins with large nuclear spins is difficult because it dramatically increases the size of the Hamiltonian and, thus, the computational cost.

We, therefore, reproduce the experimental spectra by empirically incorporating inhomogeneous linewidths.
We regard them as resulting from a static magnetic field distribution by assuming that all but the nearest-neighbor nitrogen nuclear spins are stationary during individual ODMR measurements.
Thus, the spectra are expressed by the following convolution:
\begin{equation}
I_\mathrm{conv}(B) \propto \sum_{B' = B + n\delta B} P(B'-B) I_\mathrm{sim}(B'),
\label{eq:def_Iconv}
\end{equation}
where $P(B) = \dfrac{(\Delta B/2)^2}{B^2 + (\Delta B/2)^2}$ is the shape of inhomogeneous broadening, $\Delta B = 50$~(MHz)$/\gamma_\mathrm{e}= 1.78$~mT is the effective field width of the broadening, $\delta B = 0.1$~mT, and $n=\{-25,-24,\cdots,25\}$.
We appropriately choose the range of $n$ so that the simulation reproduces the experimental spectral shape, while a slight change in $n$ does not affect the result.
We perform the normalization to ensure that the area of the spectrum remains conserved after the convolution.

The right panel of Fig.~\ref{fig:odmr_GSLAC} compares the experimental spectra $I_\mathrm{exp}$ (red markers) and the simulated spectra considering the inhomogeneous broadening $I_\mathrm{conv}$ (blue solid lines). 
They agree satisfactorily. 
It is significant to note that the agreement implies that even in GSLAC, the ODMR spectra can be explained only by the dynamics of the three adjacent $^{15}\mathrm{N}$ nuclear spins.
Slight spectral differences between the experiment and simulation would be caused by variations in experimental microwave power or by polarization or mixing of nuclear spins other than the nearest-neighbor nitrogen spins.

\subsection{\label{sec:Comparison-based-on-the-4-dip-fitting}Comparison based on the 4-dip fitting}
\begin{figure}
    \centering
    \includegraphics[width=\linewidth]{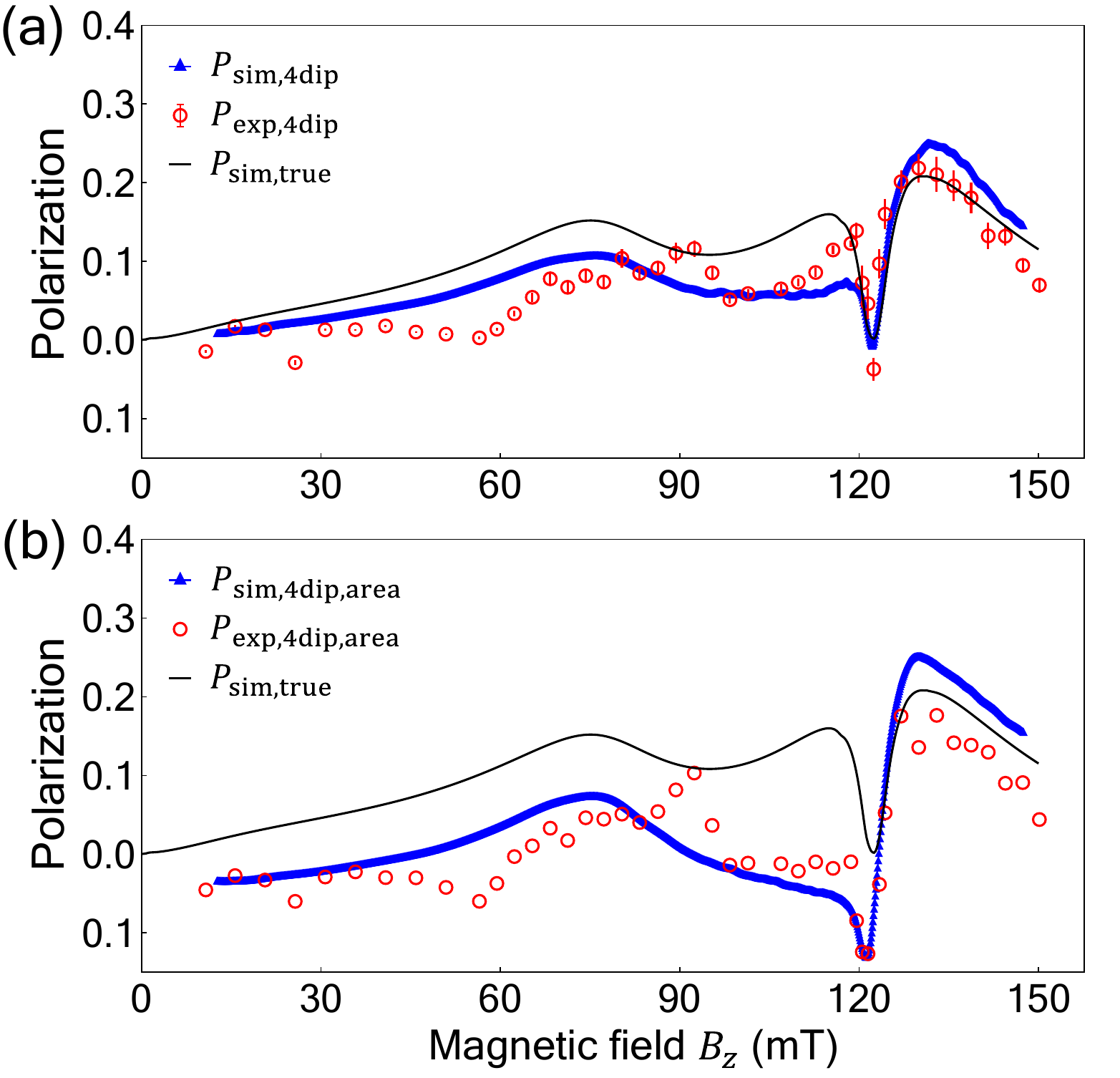}
    \caption{
    Comparison of the $^{15}\mathrm{N}$ nuclear spin polarization obtained by several methods.
    (a) Estimated polarization based on the spectral contrast.
    The blue triangles show $P_{\mathrm{sim,4dip}}$, and the red circles present $P_\mathrm{exp,4dip}$ [Eq.~\eqref{eq:def_Pol_exp}].
    (b) Estimated polarization based on the spectral area.
    The blue triangles show $P_{\mathrm{sim,4dip,area}}$, and the red circles display $P_{\mathrm{exp,4dip,area}}$ [Eq.~\eqref{eq:def_Pol_area}].
    The black solid lines in both (a) and (b) are $P_{\mathrm{sim,true}}$ [Eq.~\eqref{eq:def_Pol_sim}]. 
    $P_\mathrm{sim,true}$ and $P_\mathrm{exp,4dip}$ are the same as those shown in the top panel of Fig.~\ref{fig:pol_dep_Bz_dep_Laser}.
    }
    \label{fig:eval_accuracy_pol}
\end{figure}

We are interested in how much quantitative information about DNP can be obtained from the experimental ODMR spectra. 
Therefore, this subsection compares experiments and simulations using the 4-dip fitting. 
We treat the simulated spectra with inhomogeneous broadening as if they were experimental data and emulate the polarization estimation in the same way as in Eq.~\eqref{eq:def_Pol_exp}. 
We define the resulting polarization as $P_\mathrm{sim,4dip}$, as briefly mentioned at the end of Sec.~\ref{sec:Numerical_sim}. 

We start to investigate the validity of our model by comparing this $P_\mathrm{sim,4dip}$ and $P_\mathrm{exp,4dip}$ [Eq.~\eqref{eq:def_Pol_exp}].
Figure~\ref{fig:eval_accuracy_pol}(a) shows $P_\mathrm{sim,4dip}$ (blue triangles) and $P_\mathrm{exp,4dip}$ (red circles).
Clearly, $P_\mathrm{exp,4dip}$ and $P_\mathrm{sim,4dip}$ are in good agreement; their behavior in magnetic fields is quantitatively similar.
This agreement indicates that our simulation reproduces essential features of the experimental ODMR spectra nicely. 
We have already seen such examples in the right panel of Fig.~\ref{fig:odmr_GSLAC}.

How about the nuclear polarization?
The true polarization $P_\mathrm{sim,true}$ [Eq.~\eqref{eq:def_Pol_sim}] obtained in the simulation is superposed in Fig.~\ref{fig:eval_accuracy_pol}(a) by a solid black line.
We notice that $P_{\mathrm{sim,true}}$ and $P_\mathrm{sim,4dip}$ disagree. 
In particular, there is a difference of a few times around $B_z$=100--120~mT. 
Note that both values are purely derived from the model, not experiments.
This discrepancy implies that the accuracy of the polarization estimation by Eq.~\eqref{eq:def_Pol_exp} using the 4-dip fitting is unsatisfactory, especially around GSLAC.

Now, it is interesting to examine the conventional estimation method using the 4-dip fitting.
Usually, the following formula, which relies on the area of the spectrum instead of the contrast, is used to estimate polarization~\cite{sasaki2023nitrogen,clua2023isotopic,ru2024robust}.
\begin{equation}
P_\mathrm{exp,4dip,area} = \dfrac{\sum_{m_I} m_I A_{m_I}}{(3/2) \sum_{m_I} A_{m_I}},
\label{eq:def_Pol_area}
\end{equation}
where $A_{m_I}$ is the area of the spectral component corresponding to the $m_I$ state.
We apply Eq.~\eqref{eq:def_Pol_area} to the experimental data and the simulation. 
The resulting values are denoted as $P_\mathrm{exp,4dip,area}$ and $P_\mathrm{sim,4dip,area}$, respectively.
Figure~\ref{fig:eval_accuracy_pol}(b) compares $P_\mathrm{sim,4dip,area}$ (blue triangle markers) and $P_\mathrm{exp,4dip,area}$ (red circles) with $P_\mathrm{sim,true}$ superposed in a solid black line. 
As previously, $P_\mathrm{exp,4dip,area}$ and $P_\mathrm{sim,4dip,area}$ agree well. However, they significantly deviate from the true value $P_\mathrm{sim,true}$, especially around GSLAC.
This observation proves that the conventional area-based polarization estimation using the 4-dip fitting does not work quantitatively.

Interestingly, the results using contrast ($P_\mathrm{sim,4dip}$) are closer to the true polarization $P_\mathrm{sim,true}$ than the conventional  area-based estimation ($P_\mathrm{sim,4dip,area}$).
It may suggest the usefulness of the contrast-based fitting method for estimating polarization.

The quantitative inaccuracy of the 4-dip fitting in both experiments and simulation is due to the mixing between the $\mathrm{V_{B}^-}$ electron spin and the $^{15}\mathrm{N}$ nuclear spins. 
As the mixing becomes intense, the number of dips in $I_\mathrm{sim}$ increases, invalidating the assumptions of the 4-dip fitting, namely,  $\rho_{\mathrm{gs},m_I} \propto C_{m_I}$ or $\rho_{\mathrm{gs},m_I} \propto A_{m_I}$.
For instance, the differences between $P_\mathrm{sim,true}$ and $P_\mathrm{sim,4dip}$, as well as $P_\mathrm{sim,true}$ and $P_\mathrm{sim,4dip,area}$, are most pronounced at around GSLAC. 
The left panel of Fig.~\ref{fig:odmr_GSLAC} shows that $I_\mathrm{sim}$ contains numerous additional dips beyond the main four dips, indicating that the effects of mixing cannot be ignored. 
In contrast, at $B_z = 127.0$ mT, the edge of GSLAC, $P_\mathrm{sim,true}$ closely matches both $P_\mathrm{sim,4dip}$ and $P_\mathrm{sim,4dip,area}$ and $I_\mathrm{sim}$ exhibits a more evident four-dip structure with reduced mixing effects. 
The correlation between the number of dips in the simulated ODMR spectrum $I_\mathrm{sim}$ and the differences between $P_\mathrm{sim,true}$ and $P_\mathrm{sim,4dip}$ is consistent with the conclusion that mixing degrades the quantitative reliability of the 4-dip fitting.

\subsection{\label{subsec:supp_of_dnp} Maximum polarization}
Based on our simulation model, we discuss maximum polarization and the factors limiting it. 
Our experiments and simulations [Figs.~\ref{fig:pol_dep_Bz_dep_Laser}, \ref{fig:pol_dep_Bz_dep_mS_pm1}, and \ref{fig:eval_accuracy_pol}] suggest that the polarization of $^{15}\mathrm{N}$ nuclear spins is at most approximately 20~\%. 
A similar maximum polarization ($\sim$30~\%) was reported in the previous studies~\cite{clua2023isotopic,ru2024robust}.
These values are far from the high polarization ($\sim$100~\%) obtained in the DNP of a nitrogen nuclear spin of the NV center~\cite{jacques2009dynamic}.
Understanding the factors that limit maximum polarization is one of the most fundamental issues in DNP studies.

\begin{figure}
    \centering
    \includegraphics[width=\linewidth]{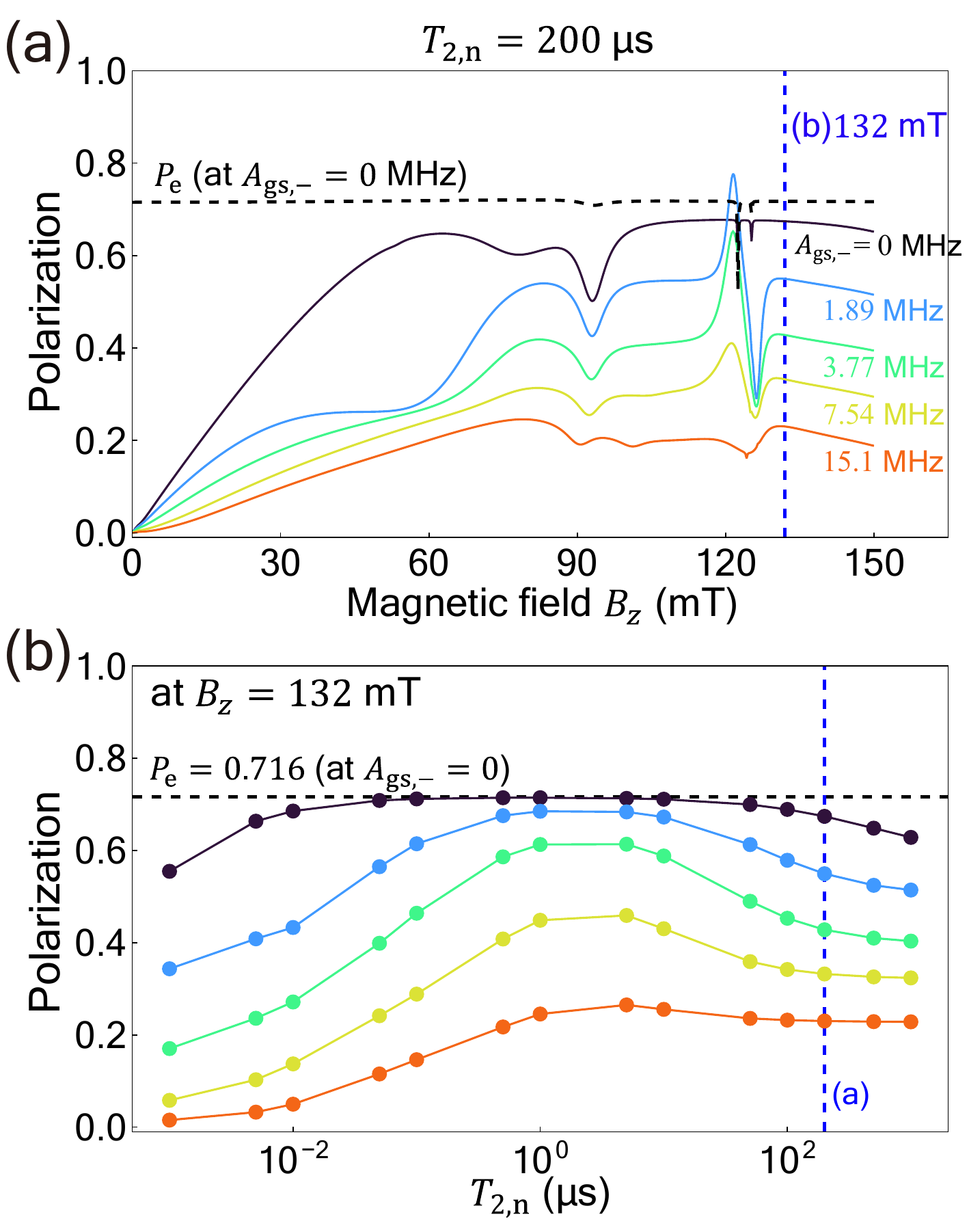}
    \caption{
    (a) $B_z$ dependence of $P_{\mathrm{sim,true}}$ for several different flip-flip interaction strengths ($A_\mathrm{gs,-}$). 
    $T_{2,\mathrm{n}}$ is fixed at 200~$\mathrm{\mu}$s. 
    (b) $T_{2,\mathrm{n}}$ dependence of $P_{\mathrm{sim,true}}$ at $B_z = 132$~mT, corresponding to (a).
    The present situation corresponds to $A_\mathrm{gs,-}=15.1$~MHz, calculated from the values in Table~\ref{tb:hyperfine_param}.
    We set $\theta = 0^{\circ}$ and $\Gamma_\mathrm{L} = 2.14$~MHz in this simulation.
    The black horizontal dashed line in (a) and (b) represents the polarization $P_{\mathrm{e}}$ between the electron spin sublevels $m_S = 0$ and $m_S = -1$ when $A_\mathrm{gs,-}=0$~MHz.
    }
    \label{fig:sim_pol_dep_flipflip_Dstate}
\end{figure}

As described in Sec.~\ref{sec:DNP_mechanism}, the flip-flip interaction $A_{-}^{(i)} \hat{S}_{+} \hat{I}_{+}^{(i)} + h.c.$ reduces the polarization~\cite{clua2023isotopic,gong2024isotope,ru2024robust}.
According to Eq.~\eqref{eq:hf_d3h}, the flip-flop and flip-flip interactions in the GS can be described by two independent parameters: $A_\mathrm{gs,+}\equiv|A_\mathrm{gs,+}^{(i)}|=|A_{\mathrm{gs},xx} + A_{\mathrm{gs},yy}|/4$ and $A_\mathrm{gs,-}\equiv|A_\mathrm{gs,-}^{(i)}|=|A_{\mathrm{gs},xx} - A_{\mathrm{gs},yy}|/4$, respectively.
To check the impact of the flip-flip term on the flip-flop term, the magnetic field dependence of the polarization $P_\mathrm{sim,true}$ simulated for several different $A_\mathrm{gs,-}$ is shown in Fig.~\ref{fig:sim_pol_dep_flipflip_Dstate}(a) while maintaining $A_\mathrm{gs,+}$. 
Except for the narrow magnetic field region ($B_z \sim121.4$~mT), the polarization increases as $A_\mathrm{gs,-}$ decreases.
If the flip-flip interaction were quenched ($A_\mathrm{gs,-}=0$~MHz), the maximum polarization could exceed 65~\%, but for the present flip-flip interaction strength, which is calculated to be $A_\mathrm{gs,-}=15.1$~MHz from the values in Table~\ref{tb:hyperfine_param}, the maximum is below 30~\%.
Thus, the flip-flip interaction is the main factor that limits the maximum polarization.

We continue to discuss in more detail to find other factors that suppress the maximum polarization.
When considering only the flip-flop interactions, the maximum polarization achievable in the DNP coincides with the electron spin polarization.
On the other hand, as indicated by the horizontal black dashed line in  Fig.~\ref{fig:sim_pol_dep_flipflip_Dstate}(a), even in the absence of the flip-flip interaction ($A_{\mathrm{gs},-} = 0$~MHz), the polarization does not match with the electron spin polarization $P_\mathrm{e}$ between the electron spin sublevels $m_S = 0$ and $m_S = -1$, as can be calculated as follows:
\begin{gather}
P_\mathrm{e} = \dfrac{\rho_{\mathrm{gs},m_S=0} - \rho_{\mathrm{gs},m_S=-1}}{\rho_{\mathrm{gs},m_S=0} + \rho_{\mathrm{gs},m_S=-1}}, \notag\\
\rho_{\mathrm{gs},m_S} = \sum_{m_I^{(i)}} \bra{\mathrm{GS}, m_S, m_I^{(i)}} \hat{\rho}_\mathrm{sat} \ket{\mathrm{GS}, m_S, m_I^{(i)}}.
\label{eq:def_Pol_e}
\end{gather}
Usually, the discrepancy between maximum nuclear spin polarization and electron spin polarization can occur owing to the fast nuclear spin relaxation.
In the simulations, we set the nuclear spin decoherence and relaxation as $1/T_{2,\mathrm{n}} = 1/200~\mathrm{\mu s}^{-1}$ and $1/T_{1,\mathrm{n}} = 1/2~\mathrm{ms}^{-1}$, respectively (Table~\ref{tb:decay_param}).
They are much smaller than the inverse of the flip-flop or flip-flip interaction strengths, so such trivial suppression should be weak.

Figure~\ref{fig:sim_pol_dep_flipflip_Dstate}(b) presents the $T_{2,\mathrm{n}}$ dependence of the polarization at 132~mT,  simulated for several different $A_\mathrm{gs,-}$ corresponding to  Fig.~\ref{fig:sim_pol_dep_flipflip_Dstate}(a).
Without the flip-flip interaction ($A_{\mathrm{gs},-}=0$~MHz), the polarization is the largest at $T_{2,\mathrm{n}} \sim 1~\mu$s, and the maximum value is nearly identical to the electron spin polarization $P_\mathrm{e}$.
This observation implies the existence of a polarization suppression that depends on nuclear spin coherence.
As expected, the polarization suppression occurs as $T_{2,\mathrm{n}}$ decreases below $\sim 0.1~\mu$s.
Interestingly, the polarization is again suppressed as $T_{2,\mathrm{n}}$ exceeds $\sim 10~\mu$s, which is counter-intuitive.
This suggests that polarization suppression for long $T_{2,\mathrm{n}}$ is not trivial simply due to spin relaxation.

Figure~\ref{fig:sim_pol_dep_flipflip_Dstate}(b) indicates that the coherence-dependent suppression appears similarly for finite $A_{\mathrm{gs},-}$. 
The maximum polarization is suppressed as the flip-flip interaction becomes stronger.
This suppression includes the influence of reduced electron spin polarization in the presence of the flip-flip interaction (data not shown).
The stronger suppression for short $T_{2,\mathrm{n}}$ and large $A_{\mathrm{gs},-}$ is due to lowering the effective polarization speed under the coexistence of flip-flop and flip-flip interactions.
The above-mentioned nontrivial suppression at long $T_{2,\mathrm{n}}$ also varies with the flip-flip interaction.
As the flip-flip interaction increases, the dependence on $T_{2,\mathrm{n}}$ weakens and converges to a constant value.
Thus, the polarization in the present interaction $A_{\mathrm{gs},-} = 15.1$~MHz remains almost constant at slightly above 20\% for $T_{2,\mathrm{n}}\gtrsim 1~\mu$s.
This behavior infers that the above nontrivial suppression for longer $T_{2,\mathrm{n}}$ is irrelevant in the present case.

\subsection{\label{subsec:role_of_dark} Role of dark states}
Finally, we examine the suppression that occurs when $T_{2,\mathrm{n}}$ is very long.
Generally, in coherent quantum systems, it is known that transitions can be suppressed by destructive interference of the entanglement states, called ``dark states''~\cite{scully1997quantum}.
The dark states tend to arise in highly symmetric systems~\cite{ruskuc2022nuclear}.
This requirement is satisfied in hBN, where the three nitrogen nuclear spins adjacent to $\mathrm{V_{B}^-}$ are equivalent.

To describe the dark state in $\mathrm{V_{B}^-}$, we represent the orthogonal bases for the three nuclear spins by the following linear combinations:
\begin{align}
\ket{\phi_1} &= \dfrac{1}{\sqrt{3}} \left( 
\ket{++-} + e^{i\frac{2\pi}{3}} \ket{+-+} + e^{i\frac{4\pi}{3}} \ket{-++}
\right), \notag\\
\ket{\phi_2} &= \dfrac{1}{\sqrt{3}} \left( 
\ket{++-} + e^{i\frac{4\pi}{3}} \ket{+-+} + e^{i\frac{2\pi}{3}} \ket{-++}
\right), \notag\\
\ket{\phi_3} &= \dfrac{1}{\sqrt{3}} \left( 
\ket{++-} + \ket{+-+} + \ket{-++}
\right), \notag\\
\ket{\pm \pm \pm} &\equiv \ket{m_I^{(1)}=\pm1/2, m_I^{(2)}=\pm1/2, m_I^{(3)}=\pm1/2}.
\label{eq:D_B_state}
\end{align}
Using these orthogonal bases, we can express the flip-flop on the $m_S=0$ in GS as follows,
\begin{align}
A_{gs,+} \sum_{i=1}^{3} \hat{S}_{-} \hat{I}_{+}^{(i)} \ket{m_S = 0} \ket{\phi_1} &= 0, \notag\\
A_{gs,+} \sum_{i=1}^{3} \hat{S}_{-} \hat{I}_{+}^{(i)} \ket{m_S = 0} \ket{\phi_2} &= 0, \notag\\
A_{gs,+} \sum_{i=1}^{3} \hat{S}_{-} \hat{I}_{+}^{(i)} \ket{m_S = 0} \ket{\phi_3} &= \sqrt{3} \ket{m_S = -1} \ket{+++}.
\label{eq:transition_D_B}
\end{align}
Since $\ket{\phi_1}$ and $\ket{\phi_2}$ are transparent to the flip-flop interaction, they are dark states.
With these dark states kept coherent, the transition to the fully polarized state $\ket{+++}$ is suppressed, reducing the achievable polarization.
This mechanism is consistent with the polarization suppression for long $T_{2,\mathrm{n}}$ observed in Fig.~\ref{fig:sim_pol_dep_flipflip_Dstate}(b).  
In other words, when these states are phase relaxed, i.e., some of them transition to $\ket{\phi_3}$, it can increase the polarization~\cite{sasaki2024suppression}.

We note that the influence of the dark state or entanglement may appear in other defects in hBN, as it is a nuclear-spin-rich crystal with high symmetry.
It would be more pronounced in isotopically enriched hBN with increased symmetry.
Despite these arguments, our simulations suggest that the impact of the dark state hardly appears experimentally in the present  case, which is attributed to the strong flip-flip interaction due to the symmetry of $\mathrm{V_{B}^-}$.
However, this does not negate the entanglement effect in other defects in hBN~\cite{Mendelson2021,chejanovsky2021coherent,GuoNatComms2023,YangACSAPN2023,scholten_multispecies_2023,Patel2023,stern2024quantum,Robertson2024,Gao2024,singh2024violetnearinfraredopticaladdressing}.
Therefore, focusing on the influence of the nuclear spin coherence in hBN remains essential.

\section{\label{sec:Conclusion}Conclusion}
We have established that the DNP and ODMR spectra in $\mathrm{h}^{10}\mathrm{B}^{15}\mathrm{N}$ can be explained by a model that considers the electron spin of $\mathrm{V_{B}^-}$ and three adjacent nitrogen nuclear spins.
Our model based on the Lindblad equation reproduced the ODMR spectra in a wide range of magnetic fields, including GSLAC.
We found that the polarization estimations based on the 4-dip fitting, including the conventional area-based method, are not highly accurate.
Developing a rigorous process to estimate the true polarization from ODMR spectra quantitatively is an important future direction.
Our model also revealed that DNP in $\mathrm{V_{B}^-}$ is limited by the flip-flip interactions and less affected by nuclear spin coherence.
We discussed the relevance of the symmetry-induced polarization suppression mechanisms, which may be observed in other defects in hBN, a nuclear-spin-rich highly-symmetric crystal.
Our quantitative understanding of the fundamental behavior of the most representative quantum defects in hBN will contribute to the theoretical developments and applications of the dynamics of electron and nuclear spins of defects in two-dimensional materials.

\begin{acknowledgments}
We thank Mr.~Tomohiko Iijima (AIST) for the usage of AIST SCR HIM for helium ion irradiations, Dr. Toshihiko Kanayama (AIST) for helpful discussions since the introduction of HIM at AIST in 2009, and Prof. Kohei M. Itoh (Keio University) for letting us use the confocal microscope system. 
This work was partially supported by
JST, CREST Grant No.~JPMJCR23I2, Japan;
Grants-in-Aid for Scientific Research (Grants No.~JP24K21194, No.~JP23K25800, and No.~JP22K03524);
“Advanced Research Infrastructure for Materials and Nanotechnology in Japan (ARIM)” (Proposal No.~JPMXP1224UT1056 and No.~JPMXP1224NM0055) of the Ministry of Education, Culture, Sports, Science and Technology of Japan (MEXT); 
“World Premier International Research Center Initiative on Materials Nanoarchitectonics (WPI-MANA)” supported by MEXT;
the Mitsubishi Foundation (Grant No.~202310021);
and the Cooperative Research Project of RIEC, Tohoku University.
Y.N. and S.N. acknowledge the financial support from FoPM, the WINGS Program, The University of Tokyo, and the JSPS Young Researcher Fellowship (No.~JP24KJ0692 and No.~JP24KJ0657). 
\end{acknowledgments}

\appendix

\section{\label{sec:Symmetry_hf}Symmetry of hyperfine interactions}
\begin{figure}
    \centering
    \includegraphics[width=\linewidth]{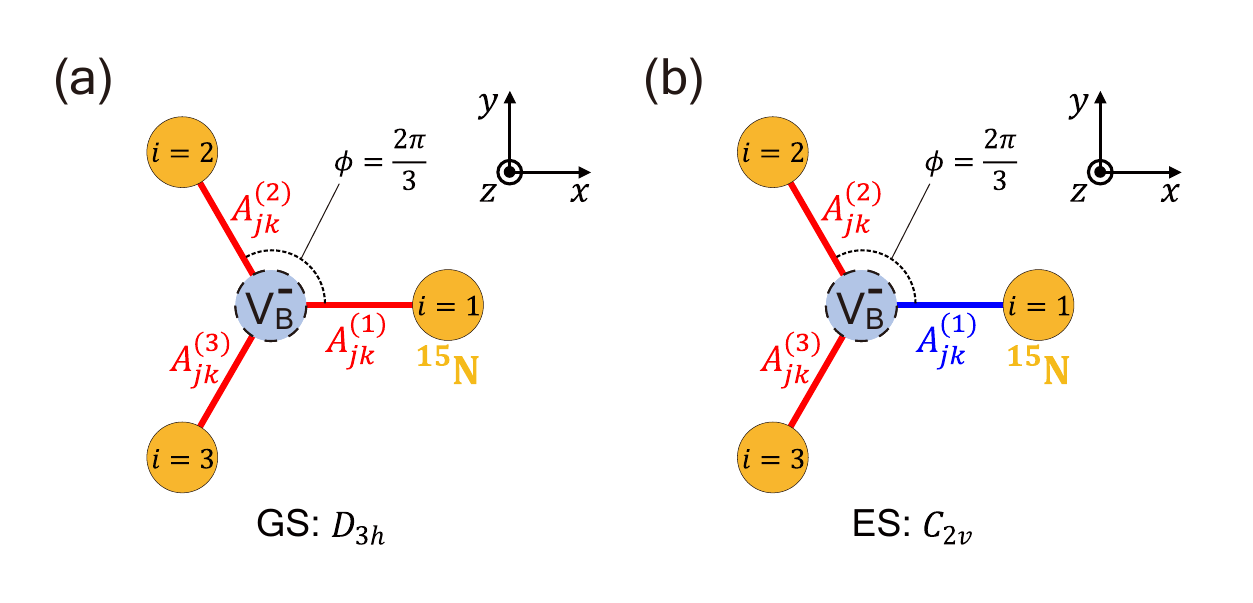}
    \caption{
    Configuration of the hyperfine interactions between the $\mathrm{V_{B}^-}$ electron spin and the three adjacent $^{15}\mathrm{N}$ nuclear spins in (a) GS  and (b) ES.
    (b) shows that, according to first-principles calculations~\cite{gao2022nuclear}, two of the hyperfine interactions (red, $i=2$ and $3$) are equivalent, while the remaining one (blue, $i=1$) is significantly smaller in the case of ES.
    }
    \label{fig:symmetry_hf}
\end{figure}

We explain the necessary conditions for the hyperfine interactions imposed by the symmetry of $\mathrm{V_{B}^{-}}$. 
Figure~\ref{fig:symmetry_hf}(a) shows a schematic of the hyperfine interactions when $\mathrm{V_{B}^{-}}$ is in the GS. 
According to first-principles calculations, the symmetry of the system in the GS is $D_{3h}$~\cite{gao2022nuclear}. 
The symmetry concerning a rotation by $\phi \equiv 2\pi/3$ in the $x$-$y$ plane requires the following condition:
\begin{gather}
A_{xx}^{(2)} = A_{xx} \mathrm{cos}^2\phi + A_{yy} \mathrm{sin}^2\phi, \notag\\
A_{yy}^{(2)} = A_{xx} \mathrm{sin}^2\phi + A_{yy} \mathrm{cos}^2\phi, \notag\\
A_{xy}^{(2)} = (A_{xx} - A_{yy}) \mathrm{sin}\phi\mathrm{cos}\phi, \notag\\
A_{yx}^{(2)} = A_{xy}^{(2)},
\label{eq:rotation_symmetry}
\end{gather}
where $A_{xx}^{(1)} \equiv A_{xx}$ and $A_{yy}^{(1)} \equiv A_{yy}$.
The mirror symmetry with respect to the $x$-$z$ plane imposes the following conditions:
\begin{gather}
A_{xy}^{(1)} = A_{yx}^{(1)} = 0, \notag\\
A_{xx}^{(3)} = A_{xx}^{(2)}, \notag\\
A_{yy}^{(3)} = A_{yy}^{(2)}, \notag\\
A_{xy}^{(3)} = -A_{xy}^{(2)}, \notag\\
A_{yx}^{(3)} = A_{xy}^{(3)}.
\label{eq:mirror_symmetry}
\end{gather}
We get Eq.~\eqref{eq:hf_d3h} from Eqs.~\eqref{eq:def_Aplus_Aminus},~\eqref{eq:rotation_symmetry},~and~\eqref{eq:mirror_symmetry}.

On the other hand, first-principles calculations indicate that in the ES, the system quickly relaxes to a state with $C_{2v}$ symmetry, as shown in Fig.~\ref{fig:symmetry_hf}(b)~\cite{gao2022nuclear}. 
From a similar discussion as in the GS case, the following relation is obtained:
\begin{gather}
A_{xy}^{(1)} = A_{yx}^{(1)} = 0, \notag\\
A_{xx}^{(2)} = A'_{xx} \mathrm{cos}^2\phi + A'_{yy} \mathrm{sin}^2\phi, \notag\\
A_{yy}^{(2)} = A'_{xx} \mathrm{sin}^2\phi + A'_{yy} \mathrm{cos}^2\phi, \notag\\
A_{xy}^{(2)} = (A'_{xx} - A'_{yy}) \mathrm{sin}\phi\mathrm{cos}\phi, \notag\\
A_{yx}^{(2)} = A_{xy}^{(2)}, \notag\\
A_{xx}^{(3)} = A_{xx}^{(2)}, \notag\\
A_{yy}^{(3)} = A_{yy}^{(2)}, \notag\\
A_{xy}^{(3)} = -A_{xy}^{(2)}, \notag\\
A_{yx}^{(3)} = A_{xy}^{(3)},
\label{eq:C2v_symmetry}
\end{gather}
where $A_{xx}^{(1)} \equiv A_{xx} \neq A'_{xx}$ and $A_{yy}^{(1)} \equiv A_{yy} \neq A'_{yy}$.
Eqs.~\eqref{eq:def_Aplus_Aminus}~and~\eqref{eq:C2v_symmetry} lead to the following conditions:
\begin{gather}
A_{\mathrm{es},+}^{(i=1)} = \dfrac{A_{\mathrm{es},xx} + A_{\mathrm{es},yy}}{4}, \notag\\
A_{\mathrm{es},+}^{(i=2,3)} = \dfrac{A'_{\mathrm{es},xx} + A'_{\mathrm{es},yy}}{4}, \notag\\
A_{\mathrm{es},-}^{(i=1)} = \dfrac{A_{\mathrm{es},xx} - A_{\mathrm{es},yy}}{4}, \notag\\
A_{\mathrm{es},-}^{(i=2,3)} = \dfrac{A'_{\mathrm{es},xx} - A'_{\mathrm{es},yy}}{4} e^{2i\phi^{(i)}}, \notag\\
\phi^{(2)} = -2\pi/3, \phi^{(3)} = -\phi^{(2)}.
\label{eq:hf_c2v}  
\end{gather}
Under $A_{xx}, A_{yy}, A'_{xx}, A'_{yy} < 0$, it immediately follows that $|A_{\mathrm{es},+}^{(i)}| > |A_{\mathrm{es},-}^{(i)}|$.

\section{\label{sec:PL_vs_Bz}Effects of magnetic field misalignment}
\begin{figure}
    \centering
    \includegraphics[width=\linewidth]{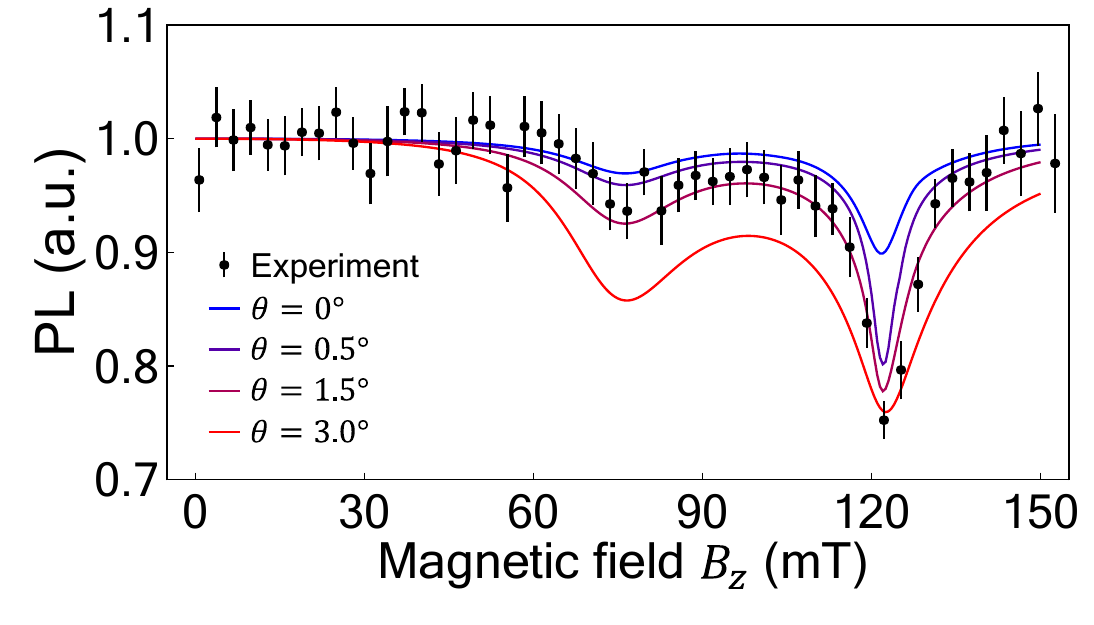}
    \caption{
    Magnetic field orientation dependence of PL intensity.
    }
    \label{fig:PL_vs_Bz}
\end{figure}

Magnetic field misalignment alters the magnetic field dependence of PL intensity.  
Figure~\ref{fig:PL_vs_Bz} shows the $B_z$ dependence of PL intensity. Black markers represent experimental data, and the solid lines in different colors correspond to PL intensity $I = \Gamma_{0} \rho_\mathrm{es}$ simulated by the steady-state solutions of the Lindblad equation without considering microwaves. Each color represents a misalignment angle $\theta = 0^{\circ}, 0.5^{\circ}, 1.5^{\circ},$ and $3.0^{\circ}$.  
We note that the experimental and simulated behaviors qualitatively agree. However, the changes in the $B_z$ dependence of PL intensity for $\theta = 0.5^{\circ}$--$1.5^{\circ}$ are not significant, making it difficult to estimate the angle $\theta$ accurately. 
The PL intensity reduction near the LAC is attributed to the mixing of the $\mathrm{V_B^-}$ electron spin ground-state $m_S$ levels~\cite{tetienne2012magnetic,yu2022excited,clua2024spin}.

\section{\label{sec:trPL}Time-resolved PL}
We experimentally estimate the optical transition parameters of $\mathrm{V_{B}^-}$ shown in Fig.~\ref{fig:intro}(b) by fitting the time-resolved PL using a five-level model of $\mathrm{V_{B}^-}$~\cite{whitefield2023magnetic, clua2024spin}. 
The estimated optical transition parameters are used in our simulations based on the Lindblad equation.

\begin{figure}
    \centering
    \includegraphics[width=\linewidth]{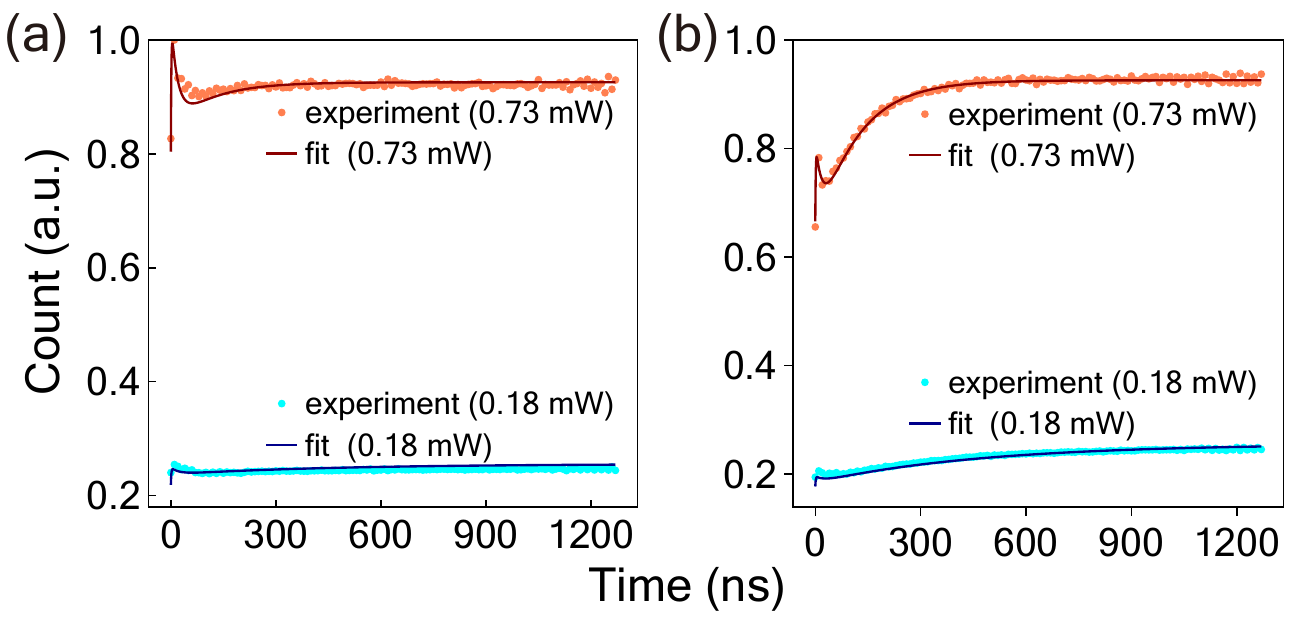}
    \caption{
    Time-resolved PL of $\mathrm{V_{B}^-}$ at $B_z=35$~mT and optical powers of 0.18 (light blue) and 0.73~mW (pink).
    The vertical axis represents photon counts, and the horizontal axis shows the time elapsed after the readout laser is turned on. 
    (a) Time-resolved PL after polarization into $m_S=0$ by laser irradiation.
    (b) Time-resolved PL after applying an adiabatic inversion pulse~\cite{spindler2016shaped,sasaki2018determination} corresponding to the $m_S=0\leftrightarrow+1$ transition.
    }
    \label{fig:time_resolved_PL}
\end{figure}

Figure~\ref{fig:time_resolved_PL} shows experimental data of the time-resolved PL obtained at $B_z=35$ mT. 
The optical powers are set at 0.18 and 0.73~mW, indicated by light blue and pink markers, respectively. 
Figure~\ref{fig:time_resolved_PL}(a) shows the time-resolved PL after polarizing to $m_S=0$ via initialization laser irradiation, and Fig.~\ref{fig:time_resolved_PL}(b) shows the time-resolved PL after applying an adiabatic inversion pulse corresponding to the $m_S=0\leftrightarrow+1$ transition~\cite{spindler2016shaped,sasaki2018determination}. 
The vertical axis represents photon counts, and the horizontal axis represents the time elapsed since the readout laser is turned on.

We fit the experimental data using the time-resolved PL described by the rate equations. 
The occupation vector for each level in Fig.~\ref{fig:intro}(b) is given by $\bm{n} = [n_\mathrm{0,gs}, n_\mathrm{1,gs}, n_\mathrm{0,es}, n_\mathrm{1,es}, n_{\mathrm{ms}}]^\mathrm{T}$, where $n_\mathrm{0,gs}$ ($n_\mathrm{0,es}$) represents the occupation of the $m_S=0$ state in the GS (ES), $n_\mathrm{1,gs}$ ($n_\mathrm{1,es}$) represents the occupation of the $m_S=\pm1$ state in the GS (ES), and $n_{\mathrm{ms}}$ corresponds to that in the MS. 
The time-resolved PL intensity at a time $t$, $\Gamma_0 (n_\mathrm{0,es} + n_\mathrm{1,es})$, follows the rate equation below:
\begin{gather}
\bm{n}(t) = e^{At}\bm{n}(t=0), \notag\\
A =\begin{pmatrix}
-\Gamma_\mathrm{L} & 0                  & \Gamma_\mathrm{0}                            & 0                                            & \gamma_{0}^{\mathrm{g}}\\
0                  & -\Gamma_\mathrm{L} & 0                                            & \Gamma_\mathrm{0}                            & \gamma_{1}^{\mathrm{g}}\\
\Gamma_\mathrm{L}  & 0                  & -(\gamma_{0}^{\mathrm{e}}+\Gamma_\mathrm{0}) & 0                                            & 0\\
0                  & \Gamma_\mathrm{L}  & 0                                            & -(\gamma_{1}^{\mathrm{e}}+\Gamma_\mathrm{0}) & 0\\
0                  & 0                  & \gamma_{0}^{\mathrm{e}}                      & \gamma_{1}^{\mathrm{e}}                      & -(\gamma_{0}^{\mathrm{g}}+\gamma_{1}^{\mathrm{g}})\\
\end{pmatrix}.
\label{eq:fit_trPL}
\end{gather}
We set the initial condition of the occupation vector in the case of Figs.~\ref{fig:time_resolved_PL}(a) and (b) as $\bm{n}(t=0)=\left[p,1-p,0,0,0\right]^\mathrm{T}$ and $\bm{n}(t=0)=\left[p-q,1-p+q,0,0,0\right]^\mathrm{T}$, respectively. 
The fitting parameters are $\{\Gamma_\mathrm{L}, \Gamma_\mathrm{0}, \gamma_{0}^{\mathrm{e}}, \gamma_{1}^{\mathrm{e}}, \gamma_{0}^{\mathrm{g}}, \gamma_{1}^{\mathrm{g}}, p, q\}$. 
The time-resolved PL obtained from this fitting is shown by the solid lines in Fig.~\ref{fig:time_resolved_PL}. 
The values estimated from this fitting are compiled in Table~\ref{tb:decay_param}.

\section{\label{sec:Lindblad_Liouville}Lindblad equation in Liouville space}
We describe the method for calculating the Lindblad equation to obtain $P_\mathrm{sim,true}$, as mentioned in Sec.~\ref{sec:Numerical_sim}. 
We vectorize the density matrix $\hat{\rho}$ as follows:
\begin{equation}
\hat{\rho} = 
\begin{pmatrix} 
  \rho_{11} & \rho_{12} & \cdots  & \rho_{1n} \\
  \rho_{21} & \rho_{22} & \cdots  & \rho_{2n} \\
  \vdots & \vdots & \ddots & \vdots \\
  \rho_{n1} & \rho_{n2} & \cdots  & \rho_{nn}
\end{pmatrix} 
\rightarrow
|\rho\rangle\rangle = 
\begin{pmatrix}\rho_{11} \\ \rho_{12} \\ \vdots \\ \rho_{1n} \\ \rho_{21} \\ \vdots \\\rho_{nn} \end{pmatrix}.
\label{eq:vectorize_rho}
\end{equation}

\hspace{20mm}

We convert Eq.~\eqref{eq:Lindblad} into the following Liouville space form~\cite{gyamfi2020fundamentals}:
\begin{gather}
\dfrac{\partial}{\partial t} |\rho\rangle\rangle = \hat{\mathscr{L}} |\rho\rangle\rangle, \notag\\
\hat{\mathscr{L}} = -\dfrac{i}{\hbar} \hat{\mathscr{H}} + \hat{\mathscr{D}}, \notag\\
\hat{\mathscr{H}} = \hat{H} \otimes I_{56} + I_{56} \otimes \hat{H}^T, \notag\\
\hat{\mathscr{D}} = \sum_{k} \Gamma_k \left\{ \hat{L}_k \otimes {\hat{L}_k}^{{\dagger}^T}
- \dfrac{1}{2} ({\hat{L}_k}^{\dagger} \hat{L}_k) \otimes I_{56} 
- \dfrac{1}{2} I_{56} \otimes ({\hat{L}_k}^{\dagger} \hat{L}_k)^T
\right\}.
\label{eq:Lindblad_Liouville}
\end{gather}
$I_{56}$ is the $56\times56$ identity matrix. 
Given that $\hat{\mathscr{L}}$ is time-independent, we calculated the steady-state solution $\hat{\rho}_\mathrm{sat}$ of Eq.~\eqref{eq:Lindblad_Liouville} by numerically solving the null space of the $3136\times3136$ matrix $\hat{\mathscr{L}}$. 
Importantly, unless the relaxation of the $^{15}\mathrm{N}$ nuclear spin is included in the Lindblad equation, the non-zero steady-state solution of Eq.~\eqref{eq:Lindblad_Liouville} is not uniquely determined.

\bibliography{reference}
\end{document}